\documentclass[12pt,preprint]{aastex}

\usepackage{mathtext,bbm,amsmath,amsfonts,amssymb,indentfirst,syntonly,graphicx}
\usepackage[english]{babel}
\usepackage{slashbox}
\usepackage{calc}
\usepackage{tikz}
\usepackage{amsmath,amssymb,amsthm}
\usepackage{latexsym,graphicx,bbm}

\newcommand{\beq}{\begin{eqnarray}}
\newcommand{\eeq}{\end{eqnarray}}

\newcommand{\p}{\partial}

\newcommand{\ba}{\left( \begin{array}}
\newcommand{\ea}{\end{array} \right)}

\shorttitle{GRB Spectra  in the complex of synchrotron and Compton processes}
\shortauthors{Y.G. Jiang}

\begin{document}

\title{GRB Spectra  in the complex radiation of synchrotron and Compton processes}

\author{Yunguo Jiang\altaffilmark{1,\ddag},   Shao-Ming Hu\altaffilmark{1,\dag}, Xu Chen\altaffilmark{1}, Kai Li\altaffilmark{1},  Di-Fu Guo\altaffilmark{1},Yu-Tong Li\altaffilmark{1}, Huai-Zhen Li\altaffilmark{2}, Hai-Nan Lin\altaffilmark{3}, Zhe Chang\altaffilmark{3,4}}
\affil{\altaffilmark{1}Shandong Provincial Key Laboratory of Optical Astronomy and \\
Solar-Terrestrial Environment,  Institute of Space Sciences, \\Shandong University at Weihai, 264209 Weihai, China}
\affil{\altaffilmark{2}Physics Department, Yuxi Normal University,  653100, Yuxi, China}
\affil{\altaffilmark{3}Institute of High Energy Physics\\Chinese Academy of Sciences, 100049 Beijing, China}
\affil{\altaffilmark{4}Theoretical Physicsl Center for Science Facilities\\Chinese Academy of Sciences, 100049 Beijing, China}
\altaffiltext{\ddag}{jiangyg@sdu.edu.cn}
\altaffiltext{\dag}{husm@sdu.edu.cn}

\begin{abstract}
Under the steady state condition, the spectrum of electrons is investigated by solving the continuity equation under the complex radiation of  both the synchrotron  and  Compton  processes.  The resulted GRB spectrum is a broken power law in both the fast and slow cooling phases. On the basis of this electron spectrum, the spectral indices of the Band function in four different phases are presented.  In the complex radiation frame, the detail investigation on physical parameters reveals that both the reverse shock photosphere model and the forward shock with strong coupling model can answer the $\alpha \sim -1$ problem. A possible marginal to fast cooling phase transition  in GRB 080916C is discussed. The time resolved spectra in different pulses of GRB 100724B, GRB 100826A and GRB 130606B  are investigated. We found that the flux is proportional to the peak energy in almost all pulses. The phases for different pulses are determined according to the spectral index revolution. We found  the strong correlations between  spectral indices and the peak energy in GRB 100826A, which can be explained by the Compton effect in the fast cooling phase. However, the complex scenario predicts a steeper index for the injected electrons, which challenges the acceleration mechanism in GRBs.
\end{abstract}


\section{Introduction \label{sec:introduction}}

Although a remarkable advance of investigations on GRBs was made in the past decades, the nature of the prompt emission of GRB is still unclear. Most spectra of GRBs can be described by the empirical Band function, whose format is a smoothly connected broken power law \citep{Band:1993,Goldstein:2012,Gruber:2014}. Many physical models have been built to explain the radiation mechanism in the prompt phase.
\citet{Beloborodov:2009be} discussed the thermal photosphere emission from the passively cooling jets. Inelastic nuclear collisions play an important role in producing the multiplicity of high energy photons \citep{Beloborodov:2003}. The resulted spectrum roughly is a broken power law except a high energy cut off, and the range of the low energy spectral index is limited.
\citet{Rees:2005} considered the leptonic dissipation model to account for the high luminosity of GRBs. One advantage of the photosphere model is the prediction of the observed peak photon energies. However, the detailed spectral indices can not be read directly from the thermal radiation to fit observations. By including the reasonable ingredients like the bulk motion of the jet, the location of the dissipation and the viewing angles, the photosphere emission can also lead to a variety of non-thermal spectral shapes \citep{Ryde:2011}.
The more popular idea about prompt emission of GRB is the relativistic fireball  model \citep{Rees:1992,Meszaros:2002,Meszaros:2006}. An extreme fast outflow  collides with the interstellar medium and produces a violent shock. Electrons in the plasma will be accelerated by the shock to form a power-law energy distribution via Fermi acceleration \citep{Peacock:1981}. The synchrotron radiation (SR) of these electrons will lead to a broken power law spectrum in the optically thin regime. The magnetic reconnection and turbulence model has been proposed by \citet{Zhang:2011} to explain the GRB emission. This model can overcome the low efficiency and electron excess problems of the internal shock model. Thanks to the great progress on the detection of high energy astrophysical photons, especially the {\it Fermi} and {\it Swift} missions, these models can be well testified by the spectral analyses of GRBs.

The {\it Fermi} Gamma-Ray Burst Monitor (GBM) spectral catalog of the first two years and four years indicated that the broken power law is still good in fitting most spectra of GRBs \citep{Goldstein:2012,Gruber:2014}. Here, we denote the low and high energy spectral indices as $\alpha$ and $\beta$, respectively. For the high quality ``BEST" samples, the low energy spectral indices $\alpha$ peak at $-1$; Up to $17\%$ samples violates the synchrotron $-2/3$ ``line of death"; while additional $18\%$ exceed the $-3/2$ synchrotron cooling limit \citep{Gruber:2014}. This challenges the standard shock plus synchrotron models. Other factors need to be considered to overcome or recoil this problem.
\citet{Lundman:2013} argued that jets with angle dependent Lorentz factor profiles can produce $\alpha \sim -1$, where photosphere emission of non-dissipative jets are considered. This suggests us that the thermal emission plus  the geometry effect  can create the observed GRB spectra.
Another remarkable model which can solve the $\alpha \sim -1$ problem is the  fast cooling synchrotron radiation in the decaying magnetic field \citep{Uhm:2014}. The low energy spectral slope $\alpha$ is found to be related to the decaying slope of the magnetic filed by an analytic method. Further on, \citet{Zhao:2014} presented the GRB spectrum with the decaying magnetic filed by considered both the synchrotron and inverse Compton (IC) radiation. The IC component is also important in generating the low energy spectral index. \citet{Duran:2012} argued that the inclusion of the IC component will harden the spectrum. But, it is still difficult for $\alpha$ to approaches the observed value $-1$, and the maximal value of $\alpha$ is obtained in an extreme condition.
In this paper, we will show that the $\alpha \sim -1$ problem can be solved in the complex of synchrotron and Compton processes, where $-1$ is not a limit but an median value of the possible range. We also investigate the time resolved spectra of several bursts, and show the evidence of the complex radiation mechanism.


\citet{Nakar:2009} discussed the synchrotron and synchrotron self-Compton (SSC) spectra with the Klein-Nishina (KN) effects, they showed that the spectral energy distribution of electrons becomes hard, and the observed synchrotron spectrum is hardened. Our work here solves the continuity equation analytically, and the spectral indices of electrons are obtained directly.
Then, we  present the full GRB spectra in the complex situation. We  show that the solution of  the $\alpha \sim -1$ problem admits quite physical parameters. The  non-trivial relations between indices and peak energies will be presented, which can be considered as a sign of the complex radiation.  The organization of the paper is arranged as follows. In Section 2, we investigate the electron spectral distribution in the complex of synchrotron and IC, which completes the analysis of \citet{Duran:2012}. In Section 3, we derive the corresponding GRB spectra in both the fast and slow cooling cases. In Section 4, we present all possible radiation phases and corresponding spectral indices. The analysis of time resolved spectra is given in Section 5. Discussions and conclusions are given in Section 6.

\section{Continuity equation of electrons}

Most works investigate the photon spectra in the prompt phase with the assumption that the injected electrons have a power-law distribution and cool down instantaneously.  The observed emissions probably originate from electrons in different emission states and regions \citep{Duran:2012}.  After interacting or radiating photons, the spectrum of electrons in the plasma changes. The electron evolution is governed by the continuity equation, which is an reduced version of the Fokker-Planck equation for electrons. The ignored terms are related to the dispersion and escaping effects.  The Larmor radius of electrons is much smaller than the emission region, and the dispersion effect is not important in the GRB scenario. In this work, we follow \citet{Duran:2012} to solve the continuity equation analytically. In the local jet frame\footnote{Here we use the prime sign to mark parameters in the jet frame.}, the continuity equation is written as
 \beq  \label{eq:continuous}
\frac{\p}{\p t} N(\gamma')+\frac{\p}{\p \gamma'}\left[\dot{\gamma'}  N(\gamma')\right]=S(\gamma'),
\eeq
where $\gamma'$ denotes the Lorentz factor of electrons. $S(\gamma')$ denotes the energy distribution of the source electrons. It is widely accepted that the shock-accelerated electrons have a power law distribution, i.e.,
\beq\label{eq:source}
S(\gamma') \propto \left\{ \begin{array}{ll}
 \gamma'^{-p}~~ &   \gamma'_{\rm m} \leq \gamma' \leq \gamma'_{\rm max}, \\
\,\,0 ~~&  \gamma' < \gamma'_{\rm m}, \quad {\rm or} \quad \gamma'>\gamma'_{\rm max}.
  \end{array} \right. \eeq
For shock accelerated electrons, $\gamma'_{\rm m}=\varepsilon_e \frac{p-2}{p-1} \frac{M_p}{M_e} \Gamma $ ($\Gamma$ is the bulk Lorentz factor of the jet, $\varepsilon_e$ is the constant fraction of the shock energy) is the minimal Lorentz factor of electrons \citep{Sari:1998}. The maximum energy for electrons gained in the shock acceleration via Fermi process  $\gamma'_{\rm max}$ is written as
$\gamma'_{\rm max} \approx \sqrt{3\pi e / \sigma_T B'} \sim 4 \times 10^{7} B'^{-1/2}$ \citep{Kumar:2012,Fan:2008}.

For a steady case, one sets $\dot{N}(\gamma')=0$. The electron will lose energy via synchrotron radiation and IC scattering processes. The cooling of the electron is described by
\beq \label{eq:coldon}
 -\dot{\gamma'}=\frac{\sigma_T B'^2 \gamma'^2}{6 \pi M_e c} + \frac{\sigma_{KN} L \gamma'^2}{4 \pi R^2 \Gamma^2 M_e c^2},
  \eeq
where $\sigma_{KN}$ is the cross section valid in the KN range.  Defining  $\eta \equiv \gamma' h\nu_{\rm seed}/M_e c^2$ ($\nu_{\rm seed}$ is the frequency of the seed photon), $\eta$ describes how deep an electron is in the KN regime. In the Thomson regime,  one has $\eta \ll 1$, while one has $\eta \geq 1$ in the KN regime \citep{Fan:2008}. Setting $\sigma_{KN}= \sigma_T f(\eta)$, $f(\eta)$ is written as \citep{Duran:2012}
\beq \label{eq:feta}
f(\eta)=\frac{3}{4}\left[ \frac{1+\eta}{\eta^3}\left(\frac{2 \eta(1+\eta)}{1+2\eta}-{\rm ln} (1+2 \eta) \right)+\frac{{\rm ln}(1+2\eta)}{2\eta}- \frac{1+3\eta}{(1+2\eta)^2} \right].
\eeq

However, one electron loses energy $ \sim \gamma' M_e c^2$ in the KN regime, and the power of Compton scattering is proportional to $  \gamma' $, rather than $\gamma'^2 $ in the second term of Equation (\ref{eq:coldon})  \citep{Fan:2005}.  We need to present a formula which is valid both in the Thomson and KN limit. Following \citet{Fan:2008}, we define a $\tilde{f}(\eta)$ function, i.e.,
\beq
\tilde{f}(\eta) \equiv  \frac{f(\eta) }{ 1+\eta}.
\eeq
The general Compton-Y parameter  is defined  as
\beq \label{eq:Y}
Y_{C} \equiv  \tilde{f}(\eta) \frac{U'_{\gamma}}{U'_{B}},
\eeq
where $U'_{\gamma}=L_{\gamma}/4\pi R^2 \Gamma^2 c$ and $U'_{B}= B'^2 /8\pi$ refer to the photon and the magnetic field energy density, respectively. $Y_C$  describes the ratio between the power of the Compton process $P'_{\rm IC}$ and that of the synchrotron radiation $P'_{\rm syn}$ in the local frame.
 Now the cooling rate of the electron is rewritten  as
\beq \label{eq:coldon2}
-\dot{\gamma'}=\frac{\gamma'^2}{T'_{\rm syn}}(1+Y_{C}).
\eeq
where  $T'_{\rm syn}\equiv 6\pi M_e c /\sigma_T B'^2$. In the complex of the synchrotron and IC processes, electrons will lose energy significantly if the cooling time $t'_{\rm c}$ is smaller than the dynamical time $t'_{\rm dyn} \approx R/\Gamma c$. When they are the same, the critical Lorentz factor of electrons is expressed as
\beq \label{eq:cri2}
\gamma'_{\rm c} \approx \frac{6\pi \Gamma M_e c^2}{\sigma_T R B'^2}\frac{1}{1+Y_{C}}.
\eeq
\citet{Duran:2012} mainly investigated the spectral slope in the low energy range. Thus, the electron spectrum was solved analytically by the continuity equation in the low energy range. We will complete the analytical solutions of the continuity equation in the full energy range. We also consider both the
fast and slow cooling phases, analogous to the full spectrum of the synchrotron radition \citep{Sari:1998,Fan:2008}.

Considering the stationary case, one sets $\dot{N}(\gamma')=0$.  The continuity equation is written as
\beq \label{eq:consyn}
\frac{\p}{\p \gamma'}\left[\dot{\gamma'}  N(\gamma')\right]=S(\gamma').
\eeq
The analytical solution is given by  $N(\gamma') \propto \dot{\gamma'}^{-1} \int S(\gamma') d \gamma'$. In the fast cooling phase, $\gamma'_{\rm c}<\gamma'<\gamma'_{\rm m}$, one has $N(\gamma') \propto \dot{\gamma'}^{-1}$. The corresponding spectral index of electrons in this energy range is
given by \citep{Duran:2012}
\beq \label{eq:p1}
p_1\equiv \left| \frac{d \,{\rm ln} N(\gamma')}{d\, {\rm ln} \gamma'}\right|=2+ \frac{d\, {\rm ln} \tilde{f}(\eta)}{d \,{\rm ln} \eta} \frac{Y_{C}}{1+Y_{C}}. \eeq
The formula of  $d \,{\rm ln} \tilde{f}(\eta)/ d \,{\rm ln} \eta $ can be approximated in the two limits, i.e.,
\beq \label{eq:f}
 \frac{d \,{\rm ln} \tilde{f} (\eta)}{d \,{\rm ln} \eta } \approx \left\{ \begin{array}{ll}
-3\eta ~~&   \eta \ll 1, \\
  -2 &   \eta \gg 1. \\
   \end{array} \right. \eeq
 One  observes that the value of $d \,{\rm ln} \tilde{f}(\eta)/ d \,{\rm ln} \eta $ is always negative. Since the Compton parameter $Y_{C}$ is always positive, the range of $p_1$ is $0 < p_1 < 2 $. This result have a significant impact on the spectral indices of GRBs.

In the slow cooling phase, i.e., $\gamma'_{\rm m}< \gamma'< \gamma'_{\rm c}$, the electrons lose no significant energy via radiation. So, the spectral distribution of electrons is roughly invariant. This leads to $N(\gamma') \propto \gamma'^{-p}$. The distribution of electrons in the high energy regime is also changed according to the continuity equation. For $\gamma' \ge {\rm max}(\gamma'_{\rm m}, \gamma'_{\rm c})$, since the source has the form $S(\gamma') \propto \gamma'^{-p}$, the corresponding power-law index can be obtained analytically
\beq \label{eq:p2}
 p_2 \equiv \left| \frac{d \,{\rm ln} N(\gamma')}{d\, {\rm ln} \gamma'}\right|=p+1+ \frac{d \,{\rm ln} \tilde{f}(\eta)}{d\, {\rm ln} \eta} \frac{Y_{C}}{1+Y_{C}}.
 \eeq
Following the argument previously, the range of $p_2$ is $(p-1)< p_2 < (p+1)$. Therefore, our work here completes the analysis given by \citet{Duran:2012}.
Collecting these results, the distribution of electrons  is written as
\beq \label{eq:spenew}
N(\gamma') \propto \left\{ \begin{array}{ll}
\gamma'^{-p_1} ~~&\gamma'_{\rm c}<\gamma'<\gamma'_{\rm m}, \\
\gamma'^{-p}~~&  \gamma'_{\rm m} < \gamma' < \gamma'_{\rm c}, \\
  \gamma'^{-p_2}~~ & {\rm max} \{\gamma'_{\rm m}, \gamma'_{\rm c} \} < \gamma' < \gamma'_{\rm max}.
  \end{array} \right. \eeq
The cutoff factor $\gamma'_{\rm max}$ is not analyzed here, since it occupies a negligible fraction. Setting $Y_{C}=0$, which means that the Compton processes are ignored, Equation (\ref{eq:spenew}) reduces to the synchrotron radiation case. From Equation (\ref{eq:spenew}), it is evident that the inclusion of Compton processes changes the spectra of electrons, and further affect the observed photon spectra.

\section{The  spectra of photons}

The  photon spectra in the prompt phase of GRB can be studied on the base of the electron spectrum in Equation (\ref{eq:spenew}).
First, we discuss the low energy synchrotron spectrum. In the fast cooling phase, the distribution of electrons is composed of two connected power laws with indices $-p_1$ and $-p_2$, respectively.
It is well known that the flux of synchrotron radiation for the low energy range is proportional to $\nu^{1/3}$. One way to estimate the flux in the low energy regime due to the synchrotron  is given by \citet{Duran:2012}, i.e.,
  \beq \label{eq:synflu}
  F^{\rm syn}_{\nu}= {\cal A}(B')\int_{\gamma'_{\nu}}^{\infty} d \gamma' N(\gamma') \left[\frac{\nu}{\nu(\gamma')} \right]^{1/3}, \eeq
where $\gamma'_{\nu}=\sqrt{2\pi M_e c \nu / e B'}$, and $\nu(\gamma')$ is the frequency corresponding to $\gamma'$. The integration $\int_{\gamma'_{\nu}}^{\infty}$ is done for $\gamma'> \gamma'_{\nu}$. If $\gamma'_{\nu}$ is smaller than $\gamma'_{c}$, then the integration is composed of tree parts, i.e., $\int^{\gamma'_{\rm c}}_{\gamma'_{\nu}}+\int^{\gamma'_{\rm m}}_{\gamma'_{\rm c}}+\int^{\gamma'_{\rm max}}_{\gamma'_{\rm m}}$. The result of the integration is independent of $\gamma'_{\nu}$, since there is no electrons distributing for $\gamma'<\gamma'_{\rm c}$. Thus,
one always has $F_{\nu} \propto \nu^{1/3}$. \citet{Duran:2012} considered that there is no lower cutoff for the electron spectrum, and obtain a formula of the spectral index. In the presence of bounded electron spectrum, their formula  is valid only for $\gamma'_{\rm c}$. We present the formula of the  spectral index at this critical frequency, i.e.,
\beq \label{eq:alpha}
 \alpha_{\rm c} \equiv \frac{d\, {\rm ln} F^{\rm syn}_{\nu_{\rm c}}}{d \, {\rm ln} \nu_{\rm c}}= \frac{1}{3}-\frac{(5/3+q)/2}{1-(p-1)/(p+2/3+q)(\gamma'_{\nu_0}/\gamma'_{\rm m})^{5/3+q}}.
 \eeq
where we define $q \equiv \frac{d \,{\rm ln} \tilde{f}(\eta)}{d\, {\rm ln} \eta} \frac{Y_{C}}{1+Y_{C}} $ for concision.
The formula here agrees with \citet{Duran:2012} except that we take use of the formula of $p_2$ in Equation (\ref{eq:p2}).  $\alpha_{\rm c}$ is meaningful to describing the spectral curvature. But $\alpha_{\rm c}$ can not describe  the full spectral slope at the turning point, since only the low energy tail of the synchrotron emission is considered to obtain $\alpha$.

When $\gamma'_{\rm c}<\gamma'_{\nu}<\gamma'_{\rm m}$, the integration contributes to the spectral index. However, Equation (\ref{eq:synflu}) counts only a small part of the synchrotron radiation power.
The photon spectrum should be derived via the relation $F_{\nu} d\nu \propto N(\gamma') P'_{\rm syn} d \gamma'$ and $\nu \propto \gamma'^{2}$ \citep{Fan:2008}. 
 By considering these relations and combining the synchrotron self-absorption, the full photon spectrum in the fast cooling phase is written as
\beq \label{eq:srfast}
F^{\rm syn}_{\nu} \propto \left\{ \begin{array}{ll}
\nu^{2}~~&   \nu < \nu_{\rm a}, \\
\nu^{\frac{1}{3}}~~&   \nu_{\rm a}< \nu < \nu_{\rm c}, \\
  \nu^{-\frac{p_1-1}{2}}~~ &   \nu_{\rm c}< \nu < \nu_{\rm m} , \\
   \nu^{-\frac{p_2-1}{2}}~~ & \nu_{\rm m} < \nu < \nu_{\rm max}.\\
       \end{array} \right. \eeq
$\nu_{a}$ refers to the absorption cut off frequency. The presence of Compton process changes the second index of the spectrum, and this breaks the ``line of death" of the index of the synchrotron radiation.

Now we discuss the spectrum in the slow lcooling case. According to Equation (\ref{eq:spenew}), the spectral indices of electrons are
$-p$ and $-p_2$ for the low and high energy range, respectively. Therefore, the synchrotron induced spectrum of photons is written as
\beq \label{eq:srslow}
F^{\rm syn}_{\nu} \propto \left\{ \begin{array}{ll}
\nu^{2}~~&   \nu < \nu_{\rm a}, \\
\nu^{\frac{1}{3}}~~&   \nu_{\rm a}< \nu < \nu_{\rm m}, \\
  \nu^{-\frac{p-1}{2}}~~ &   \nu_{\rm m}< \nu < \nu_{\rm c} , \\
   \nu^{-\frac{p_2-1}{2}}~~ & \nu_{\rm c} < \nu < \nu_{\rm max}.\\
       \end{array} \right. \eeq
Compared to Equation (\ref{eq:srfast}), we only replace $p_1$ with $p$ in the slow cooling phase. The typical value of $p$ is larger than $2$ for the shock accelerated electrons. We also have $0< p_1 <2$, see last section. Thus, the fast cooling case has a harder spectral index than the slow cooling case in the intermediate energy range. For the highest energy range, the spectrum becomes
soft. The reason is that the most energetic electrons mainly lose energy via Compton processes, and their synchrotron emission is not as efficient as the pure synchrotron radiation case.

The Compton process will produce a spectrum in the energy range above the synchrotron spectrum. The source photons of the Compton processes can be of the synchrotron origin. In this case, our scenario mainly discusses the synchrotron self-Compton (SSC) process. We estimate the energy of scattered photons in order to understand whether these photons can  be observed in GRBs. The typical bulk Lorentz factor of GRB jets in the prompt phase is several hundreds. The observed peak energy of most GRBs is about hundred keV. In the local jet frame, the energy of synchrotron photons peaks at the order of keV.  Considering the moderate shock acceleration efficiency $\epsilon \sim 0.1$, and $p=2.5$, one obtains $\gamma'_{\rm m} \sim 6000$. The $\eta$ parameter is calculated to be $\eta \sim 12$. So the SSC process happens in the KN regime. The scattering process will transit half of the electron energy to the photon. Thus, the observed photon energy can be as high as $150$ GeV, which can be recorded by the LAT monitor. However, such high energy photons can be attenuated by three significant effects. The first one  is due to  the cross section in the KN range,  which  strongly suppresses the collision chance. Secondly, the optical depth for such high energy photons can be large in the prompt phase \citep{Bosnjak:2011pt,Beloborodov:2009be,Chang:2012gq}. This will lead to the time lag phenomenons for GeV photons in GRB \cite{Abdo2009a,Abdo2009b,Ackermann2010,Ackermann2011}. The production of such  photon collision is  pairs of electron and position. This may enlarge the radiation populations in the jet slightly.  The third effect is caused by the extragalactic background lights (EBL). The absorption can be described by the model dependent gamma-ray opacity  \citep{Archambault:2014}. This effect suggests us that GRBs can enlarge the positron population in the universe.  All these effects reduce the chance of VHE photons to arrive the detector.

The power of the Compton processes in the GRB is about the same as the synchrotron, as indicated in the next section. Although the cross section suppresses the collision probability, the Compton scattering can not be ignored only by this reason. The Compton induced spectrum is most evident in blazars, which is also
produced the radiation in jets of AGNs. The spectral energy distribution (SED) of Blazars shows a bimodal pattern \citep{Bonnoli:2011}. The first peak in SED, usually in the optical to the soft X-ray range, is due to the synchrotron radiation, while the second peak in the high energy range is of the SSC or external Compton radiation. If electrons in the jet have a complex spectrum as in Equation (\ref{eq:spenew}), one can expect that the spectral indices of photons are strongly related. Since we concern the GRB spectrum in this work, the study of the complex spectrum for blazars will be given in another work \citep{Jiang:2015}

\section{ GRB spectra}

Most GRB spectra are described successfully by the empirical Band function. Now, we aim to obtain the spectral indices of the Band function by using the $F_{\nu}$ spectra in Equations (\ref{eq:srfast}) and (\ref{eq:srslow}).  The empirical Band function is a smoothly jointed broken power law, which reads
\beq \label{eq:flux}
N(E)=A \times \left\{ \begin{array}{ll}
(\frac{E}{100 \rm keV})^{\alpha} e^{-\frac{(\alpha+2)E}{E_p}},~~&   E < E_c , \\
(\frac{E}{100 \rm keV})^{\beta}e^{\beta-\alpha}(\frac{E_c}{100 \rm keV})^{\alpha-\beta} , ~~& E \geq E_c. \\
       \end{array} \right.
\eeq
$A$ is the normalization factor at $100$ keV in units of photons s$^{-1}$ cm$^{-2}$ keV$^{-1}$ \citep{Yu:2014}. The relation of $E_p$ and $E_c$ is $E_p=E_p(\alpha-\beta)/(2+\alpha)$.
Some bursts show a double broken power law spectra. In order to analyze such anomalous spectra, we introduce the third spectral index $\gamma$. Following \citet{Yu:2014}, the empirical function for the double broken power law is written as
\beq
N(E)=A \times \left\{ \begin{array}{ll}
(\frac{E}{E_n})^{\alpha} ,~~&   E < E_{b1} , \\
(\frac{E_{b1}}{E_n})^{\alpha-\beta} (\frac{E}{E_n})^{\beta}, ~~& E_{b1} \leq E < E_{b2}. \\
(\frac{E_{b1}}{E_n})^{\alpha-\beta} (\frac{E_{b2}}{E_n})^{\beta}(\frac{E}{E_n})^{\gamma},  ~~&  E \geq E_{b2}. \\
       \end{array} \right.
\eeq
Here $A$ is the normalization flux at certain energy $E_n$.

By the relation $E^2 N(E) \sim \nu F^{syn}_{\nu}$, one obtains the expression of $\alpha$, $\beta$ and $\gamma$ as functions of $p_1$ and $p_2$ in both the fast and the slow cooling phase.  There is still another phase that $\nu_c$ and $\nu_m$ is too close to distinguish, which is named as the ``marginal'' phase or the ``both" phase \citep{Yu:2014}.
Since $p_1$ and $p_2$ are functions of $p$ and $q$, one can express  $p$ and $q$ in terms of $\alpha$ and $\beta$ for a given phase. In this way, $p$ and $q$ can be  obtained from observations.

For the Band function, one remaining important question is that $E_p$ corresponds to which frequency, $\nu_m$ or $\nu_c$. Since $E_p$ is the peak energy of the $\nu F_{\nu}$ spectrum, one concrete criteria is that the spectral index of $\nu F_{\nu}$ spectrum is positive for $E<E_p$ and negative for $E>E_p$. Note that this criteria is not valid for the double broken power law.
In the fast cooling phase, $E_p$ is equal to $h\nu_m$, because $0<p_1<2$. Meanwhile, $p_2$ should be larger than $3$. This can be used as a consistent condition to constrain the physical parameters.
In the slow cooling phase,  $E_p$ is produced by electrons with $\gamma'_m$ for $p>3$. If $p<3$, $E_p$ corresponds to $h\nu_c$. So, there are two possible scenarios to explain the Band function in the slow cooling phase.
A double broken power law is possible in both the fast and slow cooling phases. Both $E_{b1}$ and $E_{b2}$ are uniquely determined if the absorption part is not included in the spectra. We will not discuss the synchrotron absorption effect unless some extreme hard low energy spectrum needs explanation.
Collecting all these results,  formulae of spectral indices are given in Table \ref{tab:SI} and \ref{tab:SI3}.

\begin{deluxetable}{ccccccc}
\tablecaption{Spectral indices in Band function \label{tab:SI}}
\tablewidth{0pt}
\tablehead{Spectral indices&\multicolumn{1}{c}{$\alpha$}&\colhead{$\beta$}&\multicolumn{1}{c}{$S$} &\multicolumn{1}{c}{$E_p$} &\multicolumn{1}{c}{$p$} &\multicolumn{1}{c}{$q$}  }
\startdata
Fast cooling  ~~& $-\frac{p_1+1}{2}$~~ & $-\frac{p_2+1}{2}$ ~~& $\frac{p-1}{2}$ ~~& $h \nu_m$ ~~&$2S+1$ ~~&$-2\alpha -3$\\
Slow cooling a  ~~& $-\frac{p+1}{2} $~~ & $-\frac{p_2+1}{2}$ ~~&$\frac{p_2-p}{2}$  ~~& $h \nu_c$~~&$-2\alpha -1$~~&$-2S -1$\\
Slow cooling b  ~~& $ -\frac{2}{3}$~~ & $-\frac{p+1}{2}$ ~~& $ \frac{p}{2}-\frac{1}{6}$ ~~& $h \nu_m$~~& $2 S +\frac{1}{3}$  & $\ldots$\\
Marginal case ~~& $  -\frac{2}{3}$ ~~ & $-\frac{p_2+1}{2}$   ~~& $\frac{p_2}{2}-\frac{1}{6}$    ~~&$h \nu_{m,c}$~~~~& $2S-\frac{2}{3}-q$ & $\ldots$\\
\enddata
\tablecomments{The parameter $S$ is defined as $S \equiv \alpha -\beta $. In the  ``Slow cooling a" case, one has $p<3$; while one has
$p>3$ in the ``Slow cooling b" case. In the marginal case, one can only determine the sum of $p$ and $q$. }
\end{deluxetable}

\begin{deluxetable}{cccccccc}
\tablecaption{Spectral indices in the double broken power law \label{tab:SI3}}
\tablewidth{0pt}
\tablehead{Spectral indices&\multicolumn{1}{c}{$\alpha$}&\colhead{$\beta$} &\colhead{$\gamma$}&\multicolumn{1}{c}{$p$}&\multicolumn{1}{c}{$q$} &\multicolumn{1}{c}{$E_{b1}$} &\multicolumn{1}{c}{$E_{b2}$} }
\startdata
Fast cooling ~~& $-\frac{2}{3}$~~& $-\frac{p_1+1}{2}$~~ & $-\frac{p_2+1}{2}$ ~~& $2(\beta-\gamma)+1$~~& $-2\beta-3$ ~~& $h\nu_c$ ~~& $h\nu_m$ \\
Slow cooling  ~~&$-\frac{2}{3}$~~& $-\frac{p+1}{2} $~~ & $-\frac{p_2+1}{2}$ ~~&$-2\beta-1$~~& $2(\beta-\gamma)-1$ ~~& $h \nu_m$ ~~& $h \nu_c$ \\
\enddata
\tablecomments{A double broken power law is possible in both the fast and slow cooling phases. $p$ denotes the spectral index of injected electrons.
$q$ denotes the correction from the Compton processes. $E_{b1}$ and $E_{b2}$ are the broken energies in the double broken power law spectrum.}
\end{deluxetable}

\section{Applications}

\subsection{The $\alpha \sim -1$  problem} \label{sec:a1}
\begin{figure}
\centering
 \plotone{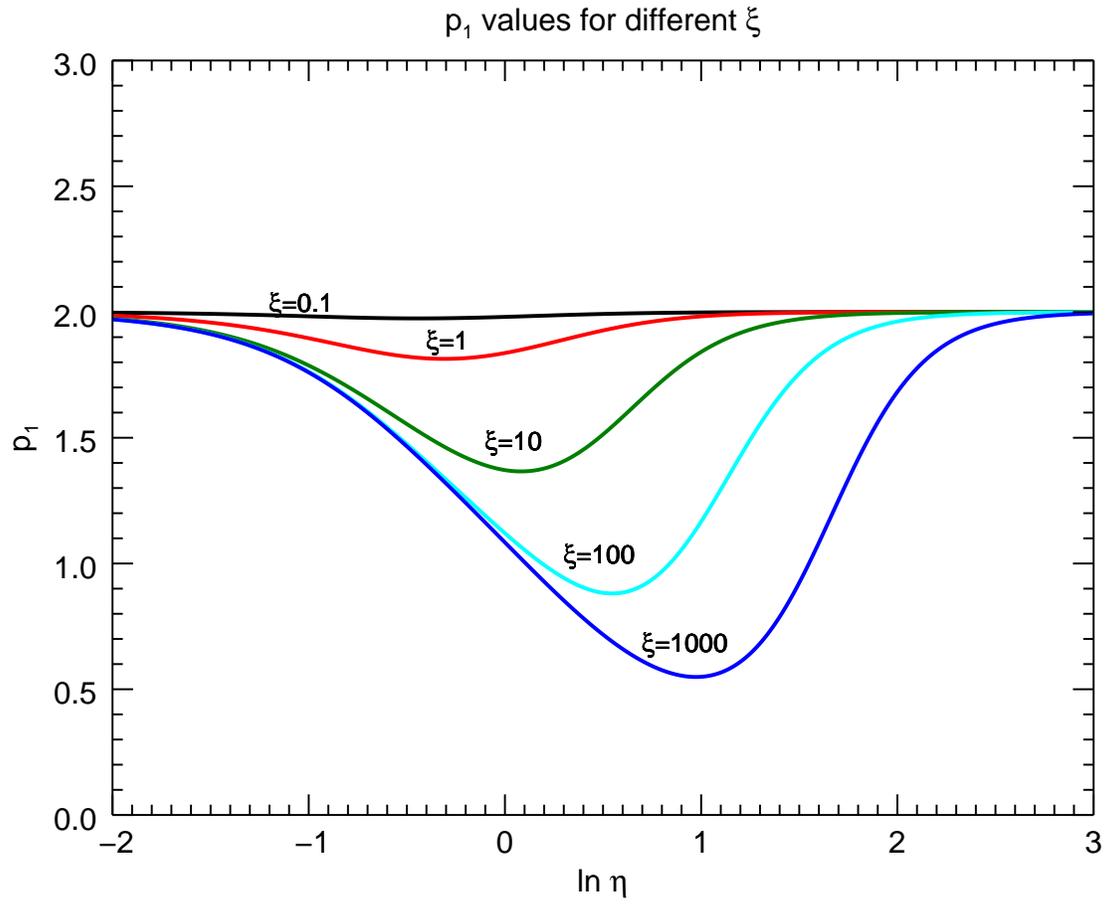}
  \caption{The index $p_1$ as a function of $\eta$ for $\xi=0.1, 1, 10, 100$ and $1000$, respectively. } \label{fig:p1}
\end{figure}

With the known expression of spectral indices, we would like to discuss the $\alpha \sim -1$ problem first. From Table \ref{tab:SI}, the fast cooling phase is the only possible case to explalin $\alpha \sim -1$. Since $p>2$, $\alpha $ in the slow cooling ``a" case is less than $-3/2$. Thus, we numerically investigate the $p_1$ value for different parameters. We define a new parameter, i.e.,
\beq \label{eq:xi1}
\xi \equiv \frac{U'_{\gamma}}{U'_B}=\frac{2L_{\gamma}}{R^2 \Gamma^2 c B'^2},
 \eeq
 which describes the ratio of the source photon energy density over the magnetic field energy density. We then plot $p_1$ as a function of $\eta$ for different values of $\xi$ in Figure \ref{fig:p1}. It is evident that the minimal value of $p_1$ decreases when $\xi$ increases.  This means that the Compton processes will strongly
flatten the electron spectrum in the low energy range. Note that the minimal value of $p_1$ can be $0$ if $\xi$ goes to infinity. Correspondingly, the
value of $\alpha$ in the band spectrum can reach $-0.5$. \citet{Duran:2012} considered  $\alpha$ in Equation (\ref{eq:alpha}), and pointed out that the extreme value of $\alpha$ is $-1$. In our analysis, $\alpha \sim -1$, or equally $p_1 \sim 1$, is not a limit but a median value in the whole parameter spaces. The range of $p_1$ can be used to fit more GRB spectra.

\begin{figure}
\centering
 \plotone{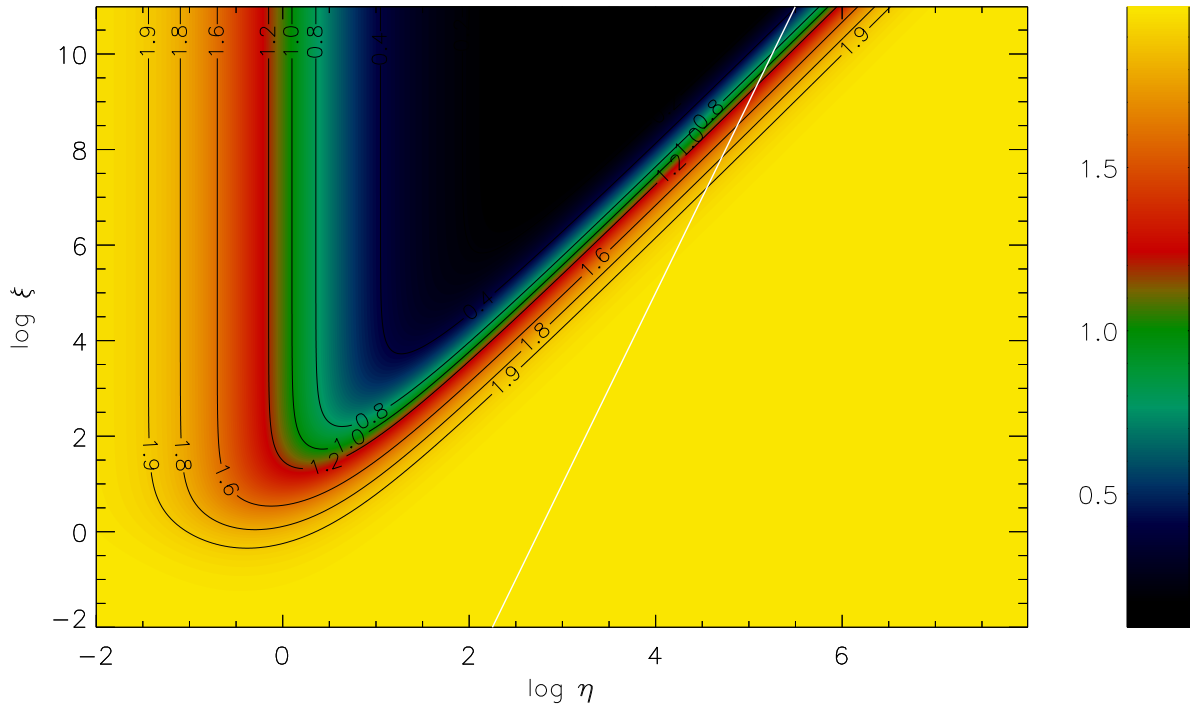}
  \caption{The contour plot of $p_1$ in the log$\eta$ and log$\xi$ plane.} \label{fig:p1all}
\end{figure}

\begin{figure}
\centering
 \plotone{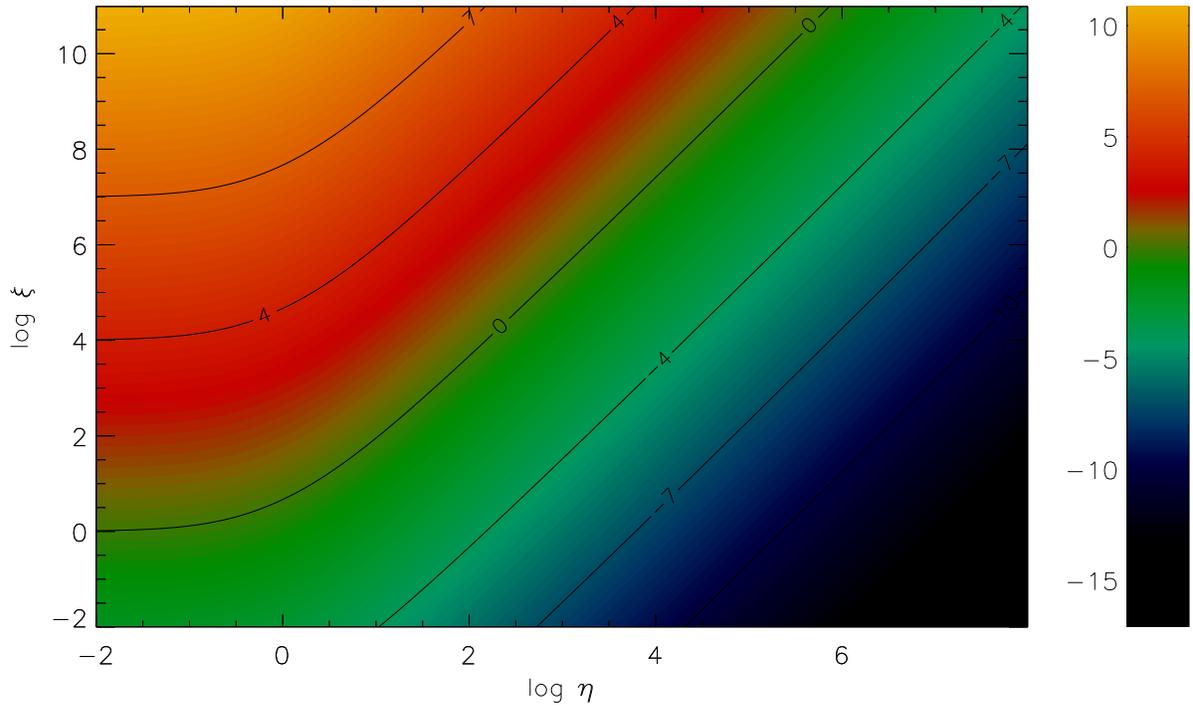}
  \caption{The contour plot of Compton $Y_C$ parameter. The numbers in the contour lines denote  values of $\log Y_C$.} \label{fig:yc}
\end{figure}

In order to show the parameter space more clearly, we plot the contour  of $p_1$ in Figure \ref{fig:p1all}, where the $x$ and $y$ axes are log$\eta$ and log$\xi$, respectively. The green region shows a ``boomerang'' pattern in Figure \ref{fig:p1all}, denoting for the $p_1 \sim 1$ zone. In the left wing of the boomerang, $\xi$ can be several tens up to $10^n$ ($n \geq4$) when $\eta$ is several. In the right wing, log$\eta$ and log$\xi$ show a linear relation. We also want to know the value of Compton parameter of $Y_C$ in the same parameter space, since it denotes which radiation mechanism is dominant. The contour plot of $Y_C$ is given in Figure \ref{fig:yc}. One can observe that  $Y_C$ is larger than 1 in  the warm  red and yellow color region, and is very small in the green and blue region. The separating zone between the warm and cold region is very close to the right wing of the boomerang, if Figure \ref{fig:yc} and \ref{fig:p1all} are overlapped. So, $p_1$ can vary significantly when  powers of the synchrotron and Compton scattering are comparable.

One question arises naturally, if these parameter values reflect the true physical conditions in GRBs. We estimate the $\eta$ and $\xi$ values for typical GRBs. $\eta$ has already been estimated to be $12$ in the last section. Here, we choose to calculate $\eta = \gamma'_p E'_p/M_e c^2$ for reference, i.e.,
 \beq  \label{eq:eta}
\eta  \approx 10 (1+z)\gamma'_{p,3} \Gamma^{-1}_2 \left(\frac{E_p}{0.5 {\rm MeV}}\right).
\eeq
The convention $Q=10^n Q_n$ is taken.  The most common observed $E_p$ is $300$ keV \citep{Goldstein:2012,Gruber:2014}. In the fast cooling phase, $\gamma'_p$ is the minimal Lorentz factor $\gamma'_m$ of injected electrons, which is a free parameter unless some specific acceleration mechanisms are specified. For the shocked electrons, one has $\gamma'_m \cong 610 \varepsilon_e \Gamma$  \citep{Sari:1998}. Thus, $\eta$ is in the order of $10$ for $\varepsilon_e \sim 0.1$.
In the synchrotron radiation, the magnetic field $B'$ can be estimated as
\beq \label{eq:B}
B' \approx 0.4 \times 10^{6} (1+z) \Gamma^{-1}_2 \gamma'^{-2}_{p,3} \left( \frac{E_p}{0.5 {\rm MeV}}\right) \,\, {\rm Gauss}.
\eeq
Once the magnetic field $B'$ is known, $\xi$ is calculated  from the definition, i.e.,
 \beq \label{eq:xi}
\xi \approx 1.1 \times 10^{-7}(1+z)^{-2} L_{52} R^{-2}_{16} \gamma'^4_{p,3}\left( \frac{E_p}{\rm 0.5MeV} \right)^{-2}.
\eeq
It seems that $\xi$  is very small. This can be realized in the magnetic dominated jet model \citep{Chang:2012}. The way to increase $\xi$ is quite limited. $R$ is constrained by the variation time of GRBs.  $E_p$ and $L_{\gamma}$ are observed variables. When we increase $\gamma'_p$, $\xi$ is significantly enlarged. But, this also increases the value of $\eta$.  One can avoid to increase $\eta$ by considering the low energy external photons, which is the case in the external Compton model. In  Figure \ref{fig:p1}, one observes that $p_1$ is smaller than $2$ unless log$\xi $ is positive.
Therefore, the left wing of the green zone is difficult to be located  in the SSC model.

In the external Compton (EC) model, by increasing $\gamma'_p$ and decreasing $E'_{\rm source}$ in the same time, one can realize that $\eta$ is unchanged while $\xi$ is greatly enhanced. When $\gamma'_p \sim 2 \times 10^5$ and $E_{\rm source} \sim 10 $ keV,  one has $\eta \sim 10$ and $\xi \sim 100$. The magnetic field is $B' \sim 10$ Gauss, which may be the shocked magnetic field in the jet. The resulted $p_1$ is in the green zone. Correspondingly, the $Y_C$ contour tells us that the power of Compton is roughly the same with that of synchrotron, which only doubles the energy budget. By this constraint, we can discard the left arm of the boomerang, where the corresponding $Y_C$ is much larger than $1$. In some bursts, a black-body (BB) component plus the Band function can better fit the observed spectra \cite{Yu:2014}. The BB bump in the spectra is found to be $10$ keV. Combining these information, the prompt emission of GRB can be explained by the reverse shock plus photosphere model. A reverse shock compresses the magnetic field and accelerates electrons in the jet, then electrons collide with the thermal photons and give synchrotron radiation in the same time. This produces the observed $E_p$ and low energy spectral index.
The high value of $\gamma'_p$ may also be due to particle acceleration in the magnetic reconnection \citep{Zhang:2002}. So, photospherical emission plus the magnetic reconnection model is a also a hopeful theory to explain the $\alpha \sim -1$ problem \citep{Zhang:2011}.

The SSC model is not excluded completely, since the right wing of  the boomerang is available. If $\gamma'_p$ increases to $10^6$, larger $\xi$ and $\eta$ can be realized.  In Equation (\ref{eq:xi}), $\xi$ depends on  $\gamma'^4_p$. One observes that the $p_1 =1$ contour is a straight line for log$\eta>2$. One needs a fine tuning of $\gamma'_p$  to land on the green zone. This strongly constrains all the relevant physical parameters.
The relation between log$\xi$ and log$\eta$ can also be described by a straight line.  One finds that ${\rm log}\xi =4 {\rm log} \eta+b$, where $b \sim -11$ denotes  all other parameters except $\gamma'_p$ in Equations (\ref{eq:eta}) and (\ref{eq:xi}).  We have marked this line with the white color in Figure \ref{fig:p1all}. The crossing point of these two lines will determine the precise value of $\gamma'_p$. We figure out roughly that $p_1$ locates at 1 when $\gamma'_p \approx 1.5 \times 10^7$. The corresponding coordinate of this point is $({\rm log}\eta, {\rm log}\xi)=(5.17,9.69)$ in Figure \ref{fig:p1}. The same point in Figure \ref{fig:yc} corresponds to $Y_C \approx 1$, which means that the power of the Compton process is the same with that of the synchrotron radiation. One also note that the $p_1\sim 1$ region and $Y_C \sim 1$ region agrees in a large area. Thus, we do not make an  energy crisis to obtain $p_1=1$.

Substituting $\gamma'_p \approx 1.5 \times 10^7$ back into Equations (\ref{eq:eta}) and (\ref{eq:B}), one obtains that $B'\approx 10^{-3}$ Gauss and $\eta \approx 1.5 \times 10^5$. The Compton scattering processes are in the deep KN regime.   In the circum-stellar medium (CSM), $B'$ is at the order of $\mu$-Gauss \citep{Kumar:2009b}. The shocked magnetic field can be amplified to hundred times larger. So, the derived  magnetic field $B'$ coincides with
external shock model. With the price of $\gamma'_p \sim 1.5 \times 10^7$, the complex radiation mechanism can be a reasonable solution to the $\alpha \sim -1$ problem. However, the extreme high value of $\gamma'_m$ challenges the  shock acceleration mechanism of electrons. The high energy electrons may be produced by the strong coupling between electrons and protons in the explosive jet, where the temperature is extremely high \citep{Zhang:2002}. Then these electrons are accelerated by the forward shock. A magnetic dominated outflow colliding with the CSM offers necessary gradients to explain all  parameters obtained here, and needs future studies.

We have used the typical parameters of GRB to calculate, but they are different from burst to burst. $\alpha$ is also different from burst to burst. The
inclusion of Compton processes offers a new method to fit the observed spectra.
Recently, it was found that $\alpha \sim -0.7$ for the time resolved spectra, which favors the synchrotron radiation model \citep{Yu:2014}. The light travel time effect, i.e. the observed photons are from electrons in different evolution stage, may also be a reason to explain $\alpha \sim -1$, since $-1$ is about the averaged value of $-1/2$ and $-3/2$. Our analysis here also indicates that $\alpha$ can be any value between
$-2/3$ and $-3/2$, if Compton processes are included.

\subsection{Phase transition}

In Table \ref{tab:SI}, we have four possible phases available to explain the GRB spectra. Phase transitions between them may occur when $\gamma'_m$ and $\gamma'_c$ vary with time. $\gamma'_m$ is determined by how electrons are injected, i.e. the acceleration processes. Except the slow cooling phase a, $\gamma'_m$ is related to the peak energy $E_{p}$. By the Equation (\ref{eq:xi1}), the peak energy $E_p$ of synchrotron origin is expresses as
\begin{align} \label{eq:ep}
E_{p}=\frac{1}{1+z} \frac{\sqrt{2}\hbar q_e}{M_e c^{3/2}}\xi^{-1/2} L_{\gamma}^{1/2}R^{-1} \gamma_{m}'^{2}.
\end{align}
Here we include the redshift to count for the cosmological effects. Obviously, one observes that $E_{p} \propto L_{\gamma}^{1/2}$, which is the Yonetoku relation \citep{Yonetoku:2004}. If we change $L_{\gamma}$ to the isotropic energy $E_{\rm iso}$, this relation is the Amati relation \citep{Amati:2002, Amati:2008}. The Amati relation is evident by statistics of many bursts \citep{Amati:2008,Yu:2014}, other parameters should be similar among different bursts.
In the fast cooling and the slow cooling b phases, one can estimate  $\gamma'_m$ by the following relation,
\begin{align} \label{eq:gamp}
\gamma'_m&=(1+z)^{1/2}(\frac{M_e c^{3/2}}{2 \hbar q_e})^{1/2} \xi^{1/4}  L_{\gamma}^{-1/4} E_p^{1/2} R^{1/2}.
\end{align}
In the ``slow cooling a'' case, $E_p$ is related to $\gamma'_c$. The critical Lorentz factor in Equation (\ref{eq:cri2}) is calculated to be
\begin{align} \label{eq:gc}
\gamma'_c & =\frac{3 \pi M_e c^3}{\sigma_T} \frac{1}{1+Y_C} \xi L_{\gamma}^{-1} \Gamma^3 R,
\end{align}
which is obtained by requiring $t'_{\rm cool} \sim t'_{\rm dyn}$. The electrons with $\gamma'_c$ radiate synchrotron photons with energy $E_c$, i.e.,
\beq \label{eq:Ec}
E_c=9\sqrt{2} \pi^2\frac{ \hbar q_e M_e c^{9/2}}{\sigma_{T}^2} \frac{1}{1+z}\big( \frac{1}{1+Y_C}\big)^2 \xi^{3/2} L_{\gamma}^{-3/2} \Gamma^6 R.
\eeq
One observes that $E_c \propto L_{\gamma}^{-3/2}$. $E_c$ strongly depends on $\Gamma$, so it can vary fast during one burst.  This relation is used to explain the anti-relation between the low energy peak and the bolometric luminosity in blazars, if the $E_c$ is the observed low peak energy \citep{Lyu:2014}.

Both $\gamma'_m$ and $\gamma'_c$ can  be changed during the prompt emission.  $\gamma'_m$ signifies the injection of the photon energy, which marks the kinematic of the jet. It can be increased in the beginning pulse of the burst. At the afterglow phase, no more energy is injected, $\gamma'_m$ will be decreased. Assuming $\gamma'_c$ is invariant for a steady state, the variation of $\gamma'_m$ leads to the phase transition. Most probably, a slow to fast transition occurs in the flux arising phase, and a fast to slow phase transition happens in the flux decaying phase.

The variation of $\gamma'_c$ depends on the radiation power. Suppose that the system is in a slow cooling phase in the beginning, which means $\gamma'_c > \gamma'_m$. And the synchrotron radiation is in a steady state. As the photon density increases (the SSC case), the Compton scattering becomes important. The Compton scattering will increase the radiation efficiency, and  $\gamma'_c$ becomes smaller, see Equation (\ref{eq:gammac}). When $\gamma'_c$ is smaller than $\gamma'_m$, a  slow to fast phase transition will occur.  Electrons with $\gamma'_{m}$ will quickly radiate away their energy, and cool down to the energy $\gamma'_{c}$. Thus, the major population of electrons will accumulate at $\gamma'_c$. Other cases of phase transition are also possible if $\gamma'_m$ and $\gamma'_c$ both vary. All these phase transitions are indicated by the change of spectral indices and the peak energy. They can be used to explain the time-resolved spectra in many bursts. With the observed $\alpha$, $\beta$ and $E_p$, one can analyze the time resolved spectra of GRBs.

 As a concrete example, we consider the time resolved spactrum of GRB 080916C. GRB 080916C is a long burst well known for its high redshift $z \sim 4.35$ and extreme luminosity $L_{\gamma} \sim 10^{54}$ erg \citep{Abdo2009a}. The minimal bulk Lorentz factor was estimated to be around $600$ to meet the optical depth constraint $\tau_{\gamma \gamma}<1$ \citep{Abdo2009a}. According to the variability time $\Delta t$, the emission radius  $R $ is estimated  to be around $\sim 10^{16} {\rm cm}$. The peak energy evolves in time. In the first time interval a ($0 \sim 3.58$ s since the trigger time), $E_p$ is about $440$ keV. It goes up to $1.1$ MeV in the second time interval ($3.58 \sim 7.68$ s), and then decreases. With these parameters, one can obtain that
 \begin{align}
 \gamma'_p& \approx 1.2\times 10^4 (1+z)^{1/2}\xi^{1/4} (\frac{E_p}{1 \rm MeV})^{1/2} L_{\gamma,54}^{-1/4} R^{1/2}_{16} ,  \label{eq:gammap}\\
 \gamma'_c & \approx 3.4 \times 10^{-3} \xi \frac{1}{1+Y_C}L^{-1}_{\gamma,54} \Gamma^3_{2} R_{16}. \label{eq:gammac}
 \end{align}
The minimal Lorentz factor $\gamma'_m$ meets the expected value of Lorentz factor for shocked electrons, i.e., $610 \varepsilon_e \Gamma \sim 0.6 \times 10^4 \varepsilon_{e,-1}\Gamma_{2}$ \citep{Sari:1998}. The critical Lorentz factor $\gamma'_c$ is much less than one. This is misleading since $\gamma'_c \geq 1$ always holds.  The derivation of $\gamma'_c$ depends on the assumption $t'_{cool} \sim t'_{dyn}$. If $t'_{cool} \ll t'_{dyn}$, $\gamma'_c$ can be much larger than one, or even larger than $\gamma'_p$.  The phase of the system is  undetermined by just comparing the derived values.

In the time interval a, $\alpha $ and $\beta$ are observed to be $-0.58\pm 0.04$ and $-2.63 \pm 0.12$, respectively. In the second time interval, $\alpha$
and $\beta$ are $-1.02 \pm 0.02$ and $ -2.21 \pm 0.03$, respectively \citep{Abdo2009a}. Then, the spectral indices do not vary significantly in the following time, and the peak energy decays.
From Table \ref{tab:SI}, if  the burst is in the fast cooling phase in the time interval a, one has $p=5.10\pm 0.32$ and $q=-1.84 \pm 0.08$. The spectral index of electrons is very large. Considering the slowing cooling b phase, one obtains that the spectral index of electrons   is $p= 4.26 \pm 0.24$.  In the marginal phase, one obtains $p+q=3.26\pm 0.24$.
If Compton process is not significant, or equally $q \approx 0$, one has $p=3.26\pm 0.24$. In the second time interval, the emission zone are most probably in the fast cooling phase. With known $\alpha $ and $\beta$, one obtains that $p = 3.38 \pm 0.1$ and $q=-0.98 \pm 0.04$. Considering the continuously electron injection, common knowledge tells us that the spectral index of injected electrons should be roughly the same in different time intervals. The slow cooling case b is not preferred for this reason.  So, the system is most probably in the marginal phase  in time interval a. With constant $p$, the variation of $q$ can account for the spectral indices in both the first and second time intervals.  In both these two phases, $E_p$ corresponds always to $\gamma'_m$. The revolution of $E_p$ agrees with that of the time resolved luminosity.  So, we have a high level confidence to claim that  a phase transition occurs in the prompt emission of GRB 080916C, i.e., transition from the marginal phase to the fast cooling phase.




\subsection{Time resolved spectra}
The time resolved spectra offers us more information on the understanding of the radiation mechanism of the prompt emission.  \citet{Yu:2014} presented
 the time resolved spectra of eight most energetic bursts with high signal-to-noise  level. The distributions of spectral indices and $E_p$ peak at $-0.73$, $-2.13$, and $374.4$ keV, respectively.  However, theses parameters are obtained from the total 299 spectra of eight bursts, the revolution of parameters in each burst was not discussed.  So, we aim to further analyse these parameters, and investigate whether the variation of spectral indices can be explained in the frame of complex radiation. We choose the data of three bursts for discussion (see Table A.2., Table A.3., and Table A.8. in \citet{Yu:2014} for reference), which are GRB 100724B, GRB 100826A, and GRB 130606B, respectively. The light curves of these three bursts are composed of simple and clear pulses, and their Band spectra have better fitting results. The fitted spectra containing parameters with only upper limits are excluded. With these selection rules, the number of the selected spectra is 114 in total. These spectra are further classified according to the pulse, we have seven pulses in total. All the pulses are illustrated in Table \ref{tab:pulse}.
We plot the $\beta \sim E_p$, $\alpha \sim E_p$, and $F_p \sim E_p$ relations for the three bursts, respectively. The peak flux $F_p$ is calculated by $F_p=E_p N(E_p)$, see Equation (\ref{eq:flux}). We use the linear relation $y=a+bx$ to fit the these relations, and results are presented in notes of figures. According to these relations, the phases of spectra will be analyzed in the following.

\begin{figure}
\centering
 \plottwo{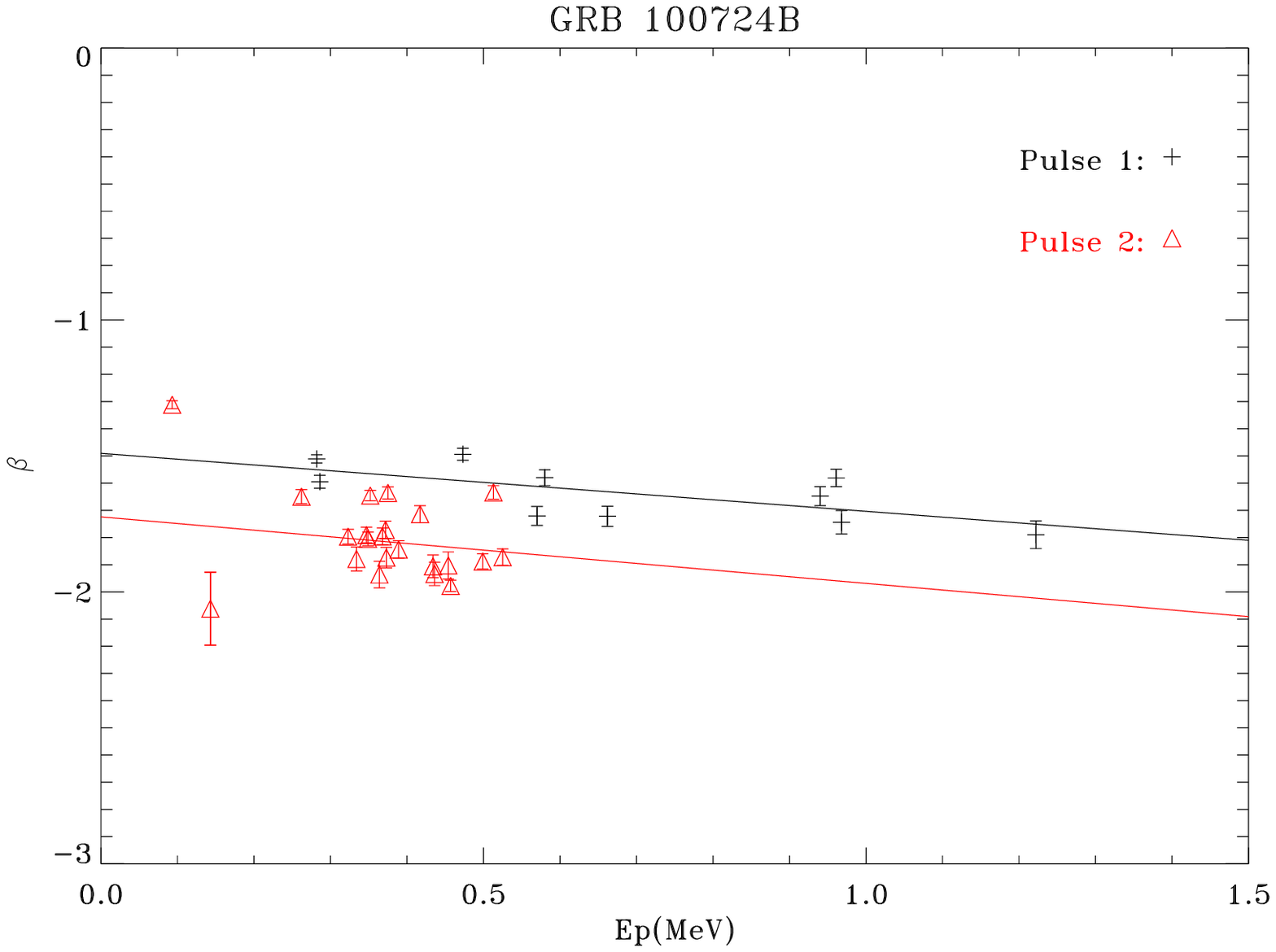}{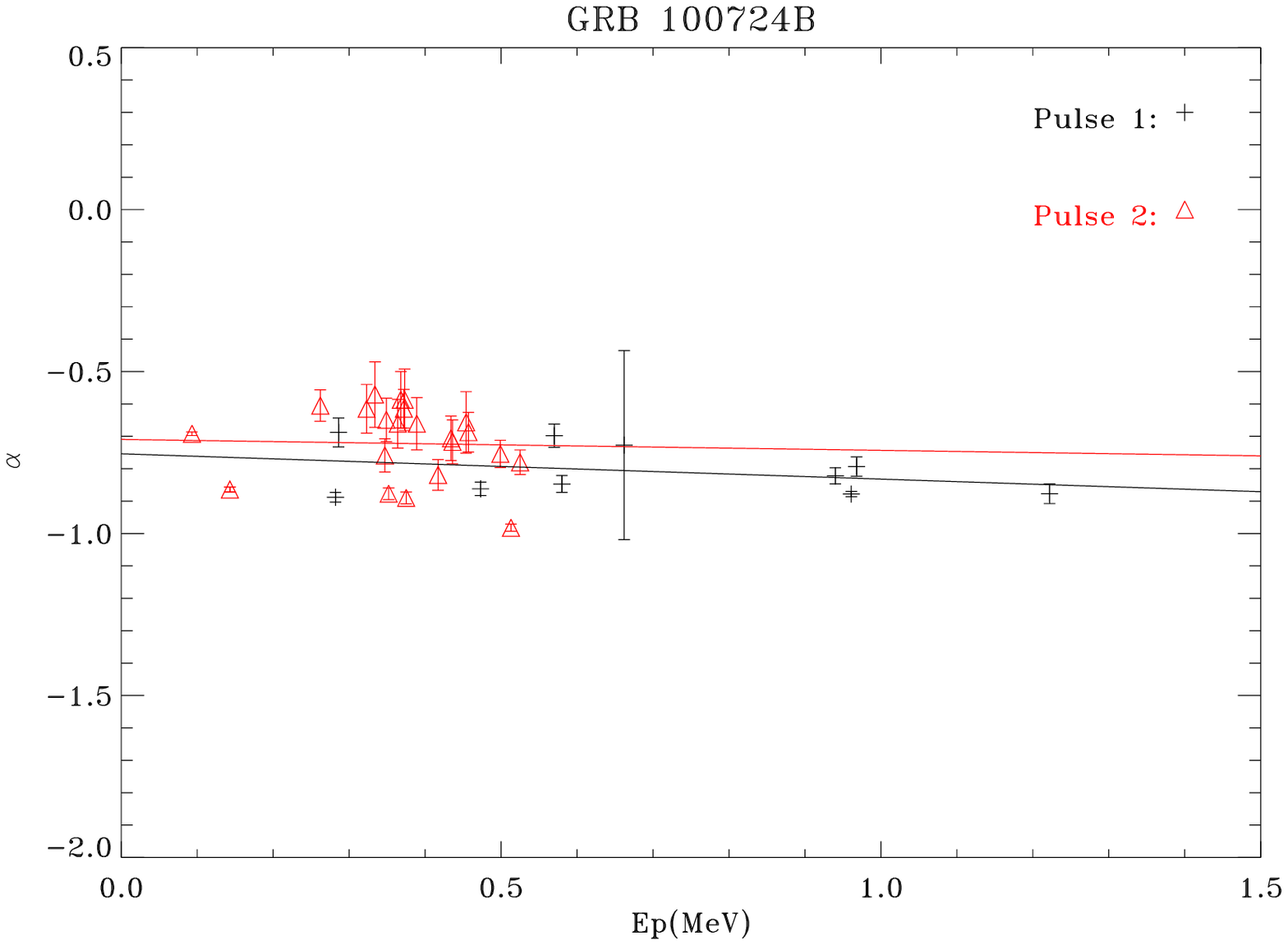}
  \caption{The plots of $\beta$ and $\alpha$ versus $ E_p$ in GRB 100724B. In the left panel, the fitting lines are described by
  $\beta=-1.491\pm0.065 -(0.213\pm 0.086)E_p$ (black line) and $\beta=-1.724\pm0.119 -(0.245\pm 0.313)E_p$ (red line) in pulse 1 and 2, respectively.
  In the right panel, the linear fittings of $\alpha$ are given by $\alpha=-0.754\pm0.062 -(0.078\pm 0.082)E_p $ (black line) and $\alpha=-0.710 \pm 0.084 -(0.034 \pm 0.222)E_p$ (red line), respectively. The value of $E_p$ is given in unit of MeV in convention.} \label{fig:724beta}
\end{figure}

GRB 100724B is composed of two pulses. Note that the second pulse is composed of several relative small peaks in the light curve, but we still consider them to be in one pulse for convenience of classification. The plots in Figure \ref{fig:724beta} indicate that $\beta$ becomes smaller when $E_p$ increases, while $\alpha$ is almost independent of $E_p$. The small slopes of two fitting lines for $\beta \sim E_p$ relation tell us that the Compton component may exist but not signify. The value of $\alpha$ also approaches $-2/3$ roughly. So, both pulses are of the marginal case. The coincidence of $h\nu_c$ and $h \nu_m$ can be reflected by the $F_p \sim E_p$ relations, which are not well linearly fitted. In the marginal phase, we can not determine $p$ and $q$ in principle, but we can obtain their constrains via the relation $p+q=-2\beta-2$. The relation of $p+q$ versus $E_p$ is plotted in Figure \ref{fig:724pq}. The linear fittings are not good for the second pulse, this probably is due to our rough classification of pulses. 

GRB 100826A contains one big pulse and one small pulse. The plot of $\beta \sim E_p$ relation indicates that $\beta$ decreases when $E_p$ increases, and also $\alpha$ have the same trend. The $F_p \sim E_p$ relation reveals that $E_p$ is proportional to the $E_p N(E_p)$ flux, which is a sign of the fast cooling phase. Thus, both pulses of GRB 100826A are in the fast cooling phase. In the complex radiation mechanism, the difference between $\alpha$ and $\beta$, i.e., $S=\alpha-\beta$ is invariant in principle. We plot the $p \sim E_p$ relation in Figure \ref{fig:100826p}, where $p$ denotes the spectral index of injected electrons. It is evident that $p$ remains almost invariant when $E_p$ varies. The $q \sim E_p$ relation is also plotted in Figure \ref{fig:100826p}. Except some exceptional points, there is a significant correlation between $q$ and $E_p$. The intercepts of the fitting lines are near $-2$. The two lines have  different slopes, which indicates the intrinsic physical parameters are different for the two pulses. The correlation between $q$ and $E_p$ indicates that the Compton scattering is in the deep KN regime. The different trend of $p$ and $q$ dependence on $E_p$ can be considered as an evidence for the complex radiation mechanism.

The light curves in GRB 130606B, see Fig.A.2. in \citep{Yu:2014}, have four pulses in total. In Table \ref{tab:pulse}, we combined the latter two pulses as one. This may bring some large errors in linear fitting. In Figure \ref{fig:130606ba}, $\beta$ and $\alpha$ have different dependence on $E_p$ for three pulses. In pulse 1, the $\alpha$ and $\beta$ are independent of $E_p$, and $\alpha$ approaches $-2/3$. The $F\sim E_p$ relation  illustrates that pulse 1 is in the slow cooling b or  marginal case. In the slow cooling b case, the injected electrons should have a spectral index $p > 3$, this phase can not be excluded. The nearly zero slope of the linear fitting of $\beta \sim E_p$ relation indicates that the synchrotron radiation is dominant. In pulse b, both $\beta$ and $\alpha$ show a slightly increasing trend  when $E_p$ increases. However, the linear fitting has a large error. The most possible phase in pulse 2 is the fast cooling with the pure synchrotron radiation, since the $F \sim E_p$ relation shows a nice linear fitting, see Figure \ref{fig:130606flux}. Pulse 3 have a large error correction for the $\beta \sim E_p$ fitting, while $\alpha$ has the similar behavior with pulse 1. The $F \sim E_p $ relation indicates a good linear fitting with slope $1.636\pm0.229$. So, both slow cooling b and marginal case are possible in pulse 3.

The time resolved spectra show that  prompt emissions of GRBs have variable phases from pulse to pulse. The $F_p \sim E_p$ relations indicate that $E_p$ mostly corresponds to $h\nu_m$, the slow cooling a phase is not evident in these three bursts. The spectra of GRB 100826A contains the feature of complex emission in a high significant manner. Such kind of feature is less evident in GRB 100724B and GRB 130606B. \citet{Frontera:2012} found the correlation between $\alpha$ and $E_p$  in GRB 980329, which is also a burst with single one pulse. One  also observes that  the flux rising phase and flux decaying phase have no differences in these relations during one pulse. This hints us that it is a better method to classify the time resolved spectra  according to pulses. The $F_p\sim E_p$ relations in their work also show  the nice linear fittings in GRB 970111, GRB 980329, GRB 990123, and GRB 990510 \citep{Frontera:2012}. Our results agree with theirs. The marginal phase seems to be popular in the bursts. In this case, $\gamma'_m$ and $\gamma'_c$ is close to each other. \citet{Yu:2014} also revealed that there is a universal break ratio between $\nu_m$ and $\nu_c$, which is less than 10.  Since $\gamma'_m$ corresponds to the acceleration processes, while $\gamma'_c$ is related to the emission power of electrons, the agreement of them indicates the balance between the the acceleration and radiation processes. This is reflected by the fact that the Band function is suitable to describe the spectra in different time slices during one burst.

\begin{deluxetable}{ccccc}
\tablecaption{The pulses of GRBs \label{tab:pulse}}
\tablewidth{0pt}
\tablehead{GRB&\multicolumn{1}{c}{Pulse}&\colhead{Time interval (s)}&\multicolumn{1}{c}{Number of spectra} &\multicolumn{1}{c}{Phases} }
\startdata
100724B  ~~& 1 ~~ & $-7.186 \sim 41.089 $ ~~& 10~~& Marginal \\
\phantom{100724B} ~~& 2 ~~ & $41.089 \sim 130.458 ~~$ ~~& 23 ~~& Marginal ~~\\
100826A  ~~& 1 ~~ & $-2.048 \sim 40.792 $ ~~& 36~~& Fast cooling\\
\phantom{100826B} ~~& 2 ~~ & $40.792 \sim 98.549 ~~$ ~~& 14 ~~& Fast cooling\\
130606B  ~~& 1 ~~ & $-3.072 \sim 11.459 $ ~~& 7~~& Slow cooling b/Marginal \\
\phantom{130606B} ~~& 2 ~~ & $11.459 \sim 26.218 ~~$ ~~& 11 ~~& Fast cooling \\
\phantom{130606B} ~~& 3 ~~ & $26.218 \sim 77.824 ~~$ ~~& 13~~& Slow cooling b/Marginal \\
\enddata
\tablecomments{GRB pulses and their phases}
\end{deluxetable}

\section{Discussion and conclusion}

By considering both the SR and the  Compton  processes, the continuity equation of electrons is investigated.  We find that the spectrum of electrons is a broken power law in both the fast and slow cooling phases. The inclusion of the Compton processes changes the spectral indices of photons. Analytical expression of spectral indices in four different phases of GRBs is given, which enriches the theoretical implications of the Band function. The $\alpha \sim -1$ problem can be solved in this new complex radiation frame. A detailed investigation of physical parameters shed lights on the prompt emissions of GRBs. In the EC case, the reverse shock plus the photosphere model is favored to explain the prompt emission. The nearly equal powers of Compton and synchrotron radiations do not overburden the energy budget. Also, the SSC case can account for the spectral index and peak energy, if forward shock and
strong coupling between electrons and heavy ions are considered. These arguments indicate that the popular shock models of GRBs can also answer the low energy spectral index problem.

Phase transitions are also possible during the prompt emission of one GRB. We show that GRB 100826A experiences the phase transition, from the marginal phase to the fast cooling phase. We analyse the time resolved spectra of three bursts, which are GRB 100724B, GRB 100826A and GRB 130606B, respectively. The plots of  indices versus peak energy show many interesting features of the prompt emission. Different pulses have different correlation between indices and peak energies. Even in one burst, pulses can be in the different radiation phases. The correlations of $\beta \sim E_p$ and $\alpha \sim E_p$ shows the signature of the complex radiation in the fast cooling phase, especially in GRB 100826A. The pure synchrotron radiation can not explain the revolution of spectral indices. \citet{Frontera:2012} has reported the dependence of $\alpha$ on $E_p$ in GRB 980329, which is also of one pulse burst. We guess that the spectral indices dependence on the peak energy can be found in the pulse dependent spectra. \citet{Titarchuk:2012} argued that the GRB spectra are formed two upscattering processes. The resulted spectral index of Comptonization has  relations with the bulk motion of the outflow and the otical depth. Such model also predicts a correlation between $E_p$ and $L_{iso}$ for time resolved spectra\citep{Frontera:2012}. However, $\beta$ and $\alpha$ are not related in the Comptonization theory. This can be considered as a special feature of the complex radiation in the fast cooling phase.

In the complex radiation scenario, the derived $p$ for the injected electrons is larger than 3, e.g., $p\approx 3.6$ in GRB 100826A.  This challenges the widely accepted electron acceleration mechanism. For shock acceleration, the typical value of $p$ is from $2.3$ to $2.8$ in the magnetized collisionless pair shock \citep{Sironi:2009}. \citet{Guo:2014} have found that a hard power law of electrons can be formed in the relativistic magnetic reconnection, and $p$ can be as hard as $1$. The analytical way to obtain this hard spectra is also solved by the continuity equation in the unsteady state, where the injection, acceleration and escaping terms are considered. It was shown that the first Fermi acceleration can not produce a power law spectra without the injection \citep{Guo:2014}. These numerical results are not verified in observation, especially from GRB. The conflict here is a two sides thing. One side is that the soft injected spectrum questions the shock acceleration paradigm of GRB, specific physical configurations are needed to explain the steeper spectral index. Even in the pure synchrotron radiation, the derived $p$ from Band fitting does not agree with the shock acceleration completely.
The other side is that our analysis is questioned. In our derivation of the continuity equation, we consider only the single power law injection. If the injected electrons have a broken power law spectrum, the resulted photon spectrum is extended to double broken power law. Then, the Band function may be fitted in the shock acceleration model. The more complicate spectrum can be obtained by complicating the injected spectrum.
We did not consider the acceleration term and the time evolution term. If these terms are included, the power law solution still exists from analytical argument, and other corrections to the index can be expected.

In our work, the Compton induced spectra are not discussed. Roughly, the spectral index in the high energy range induced by Compton processes are $p_2$, since the electrons share their energy with photons. The observation of the Compton induced spectrum is difficult in GRBs, see Section 3. One can expect that this high energy spectral index depends on the synchrotron peak energy and the IC peak energy. The Compton spectra are most evident in blazars. \citet{Abdo:2010b} found a harder-brighter tendency in the gamma ray band for certain specific objects of the subclasses of blazars, i.e., FSRQs and  LBLs. \citet{Lyu:2014} indicated that FSRQs and LBLs are in the slow cooling phase, because there is an anti-correlation between luminosity and peak energy. Combining them together, one can infer that the spectral index in the gamma ray band becomes softer when the peak energy increases for FSRQs and LBLs. For a sample of bright blazars, a correlation between the spectral index and the peak energy was found \citep{Abdo:2010}. The linear fitting of the correlation is ${\rm log}\nu^{IC}_{p} =-4 \Gamma+31.6 $ ($\Gamma$ is the Fermi spectral index), see Figure 29 in \citep{Abdo:2010}.  The anti-correlation is against with that of GRB 100826A, which is obtained from the time resolved spectra. The disagreement does not make a real conflict, since blazars can be in the slow cooling phase, while GRBs are in the fast cooling phase. The investigation on the Compton induced spectral indices will be given in the future \citep{Jiang:2015}.

One can conclude that the correlation between indices and peak energy (or luminosity) exists in both the time resolved and time integrated spectra, and this applies to both GRBs and blazars. Other correlations between spectral index and the peak energy are possible, see the time resolved spectra of the GRB 130606B. The spectral revolution is one main reason to cause the differences between the spectra of peak flux and  the time integrated spectra, and a large number of GRBs can have such behavior (see Figure 17 in \citet{Goldstein:2012}).  Our analysis of the time resolved spectra in this work show the pulse dependent spectral revolution, and proposes a new manner to reveal the  prompt emissions of GRBs. The richness of the complex radiation mechanism  enables us to explain many observed spectral phenomena in astrophysical objects, especially the the GRBs and the blazars.

\begin{figure}
\centering
 \plotone{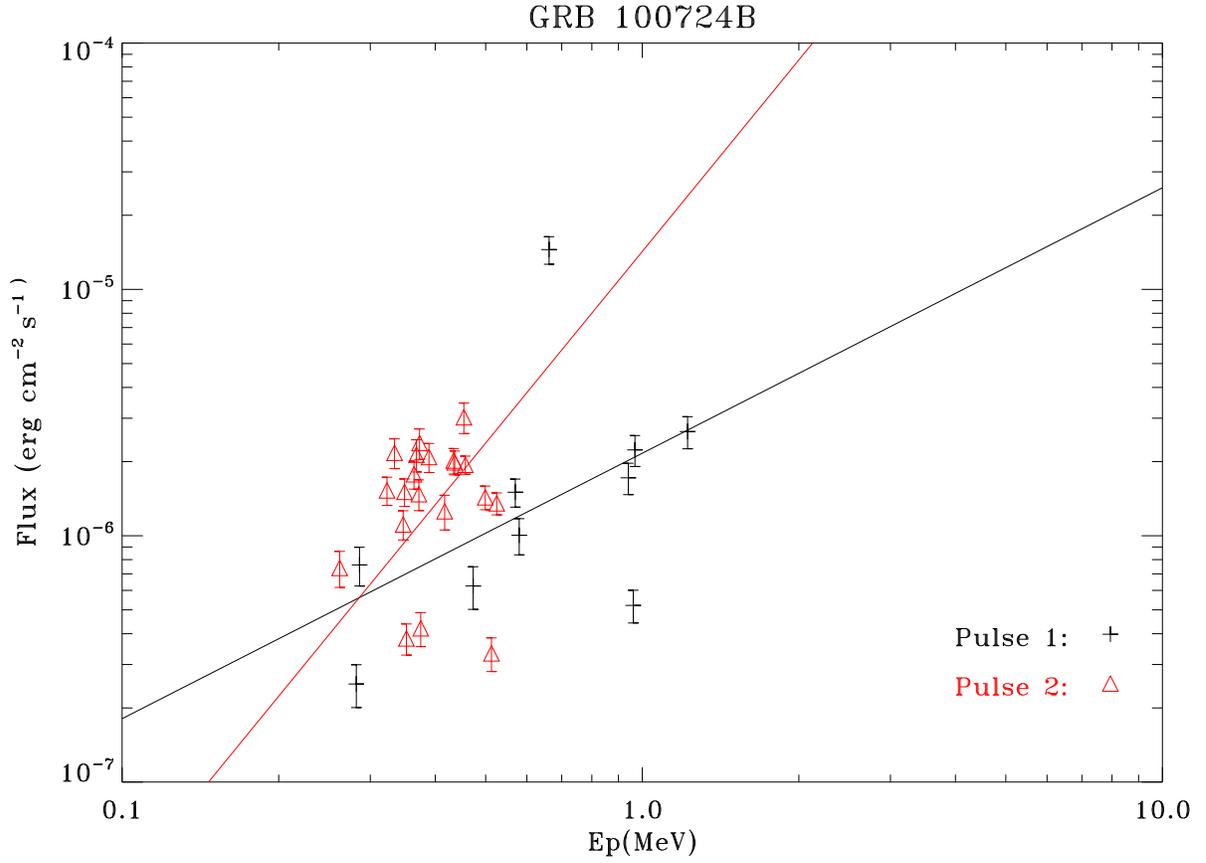}
  \caption{The  plot of $ F_p \sim E_p$ relation in GRB 100724B. The fitting lines are given by ${\rm log}F_p=-5.665 \pm 0.197 +(1.078\pm0.673){\rm log}E_p$ (black) and ${\rm log}F_p=-4.846 \pm 0.181 +(2.583\pm0.356){\rm log}E_p$ (red), respectively. } \label{fig:724flux}
\end{figure}

\begin{figure}
\centering
 \plotone{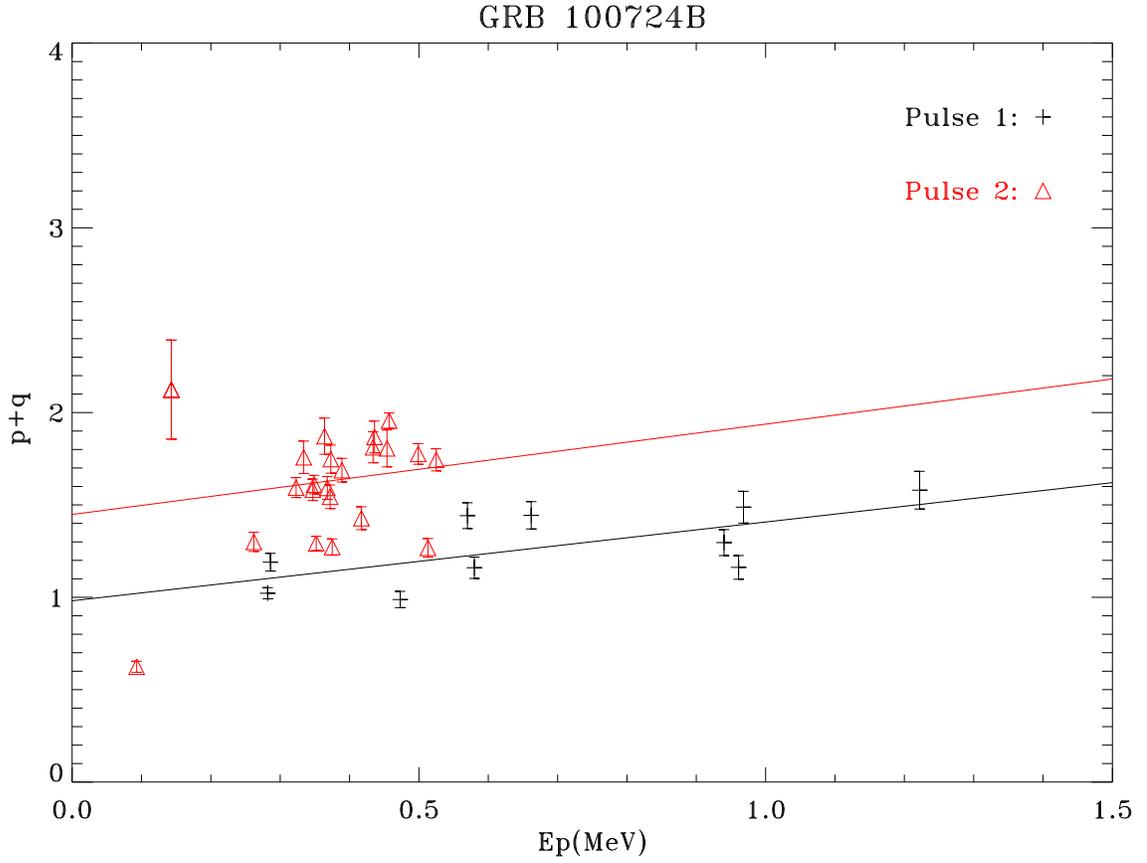}
  \caption{The plot of $p+q$ versus $E_p$ relation in GRB 100724B. In pulse one, the linear fitting is given by $p+q=0.981\pm 0.130+(0.426\pm0.171)E_p$,
   In pulse two, the fitting line is given by $p+q=1.448\pm0.237 +(0.489\pm0.627)E_p$.} \label{fig:724pq}
\end{figure}

\begin{figure}
\centering
 \plottwo{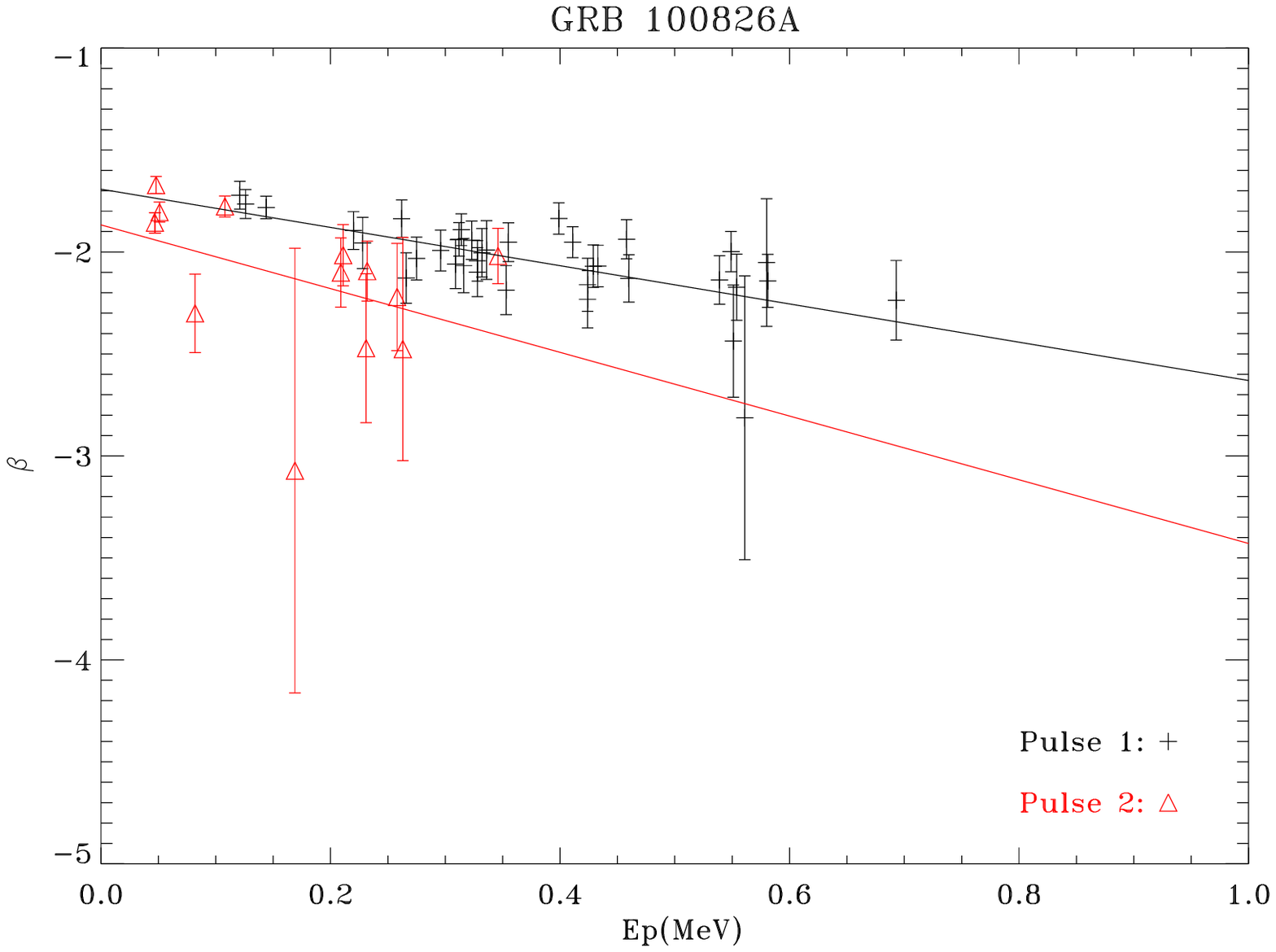}{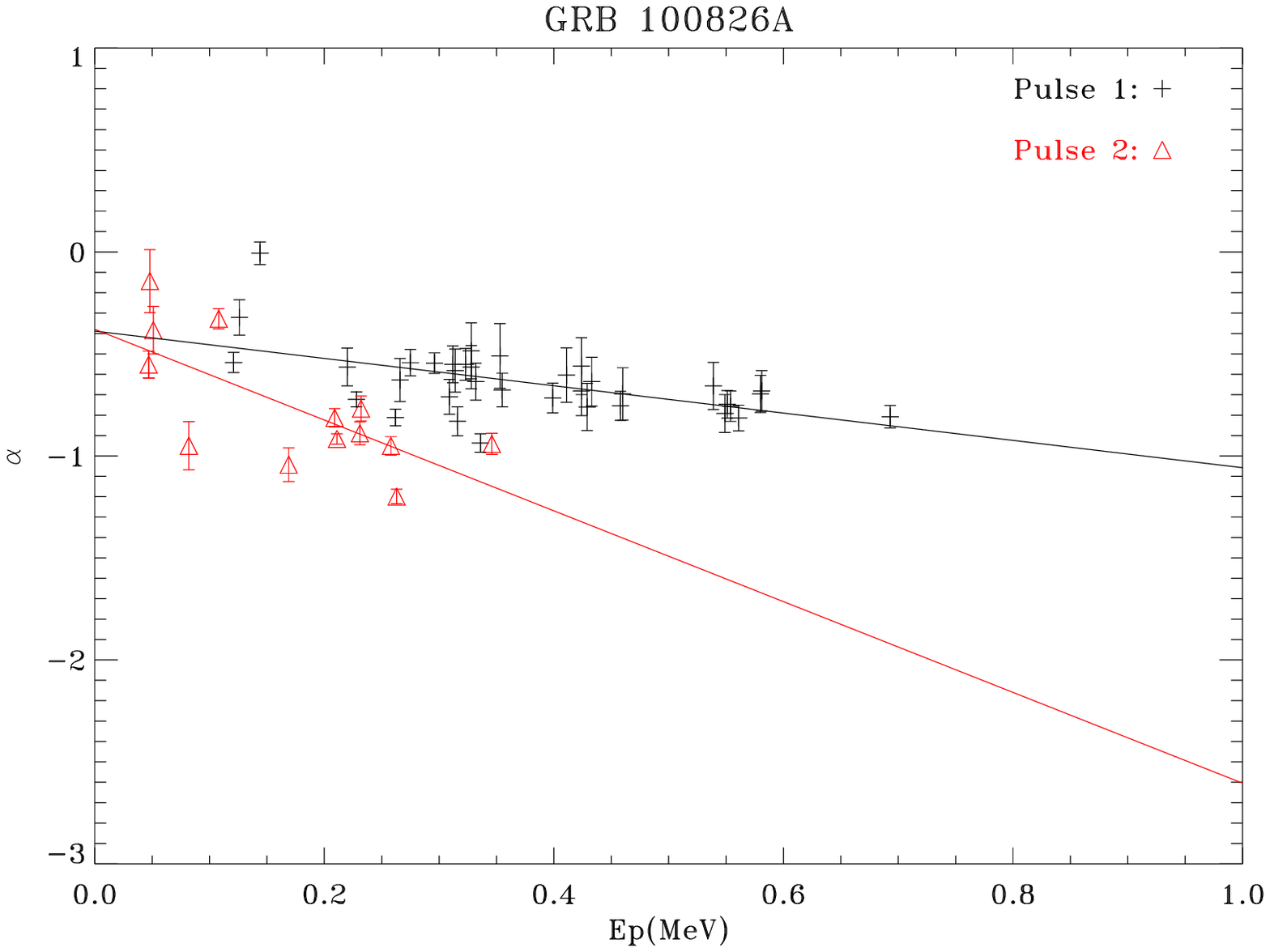}
  \caption{The plot of $\beta$ and $\alpha$ versus $E_p$ relation in GRB 100826A. In the left panel, the fitting lines are described
  by $\beta=-1.692\pm0.074-(0.938\pm0.186)E_p$ (black) and $\beta=-1.868\pm0.185-(1.562\pm0.974)E_p$ (red), respectively. In the left panel,
  the linear fittings of $\alpha$ are given by $\alpha=-0.388\pm0.067 -(0.670 \pm0.168) E_p$ (black) and $\alpha=-0.379\pm0.116 -(2.223 \pm0.612) E_p$ (red), respectively.   } \label{fig:100826beta}
\end{figure}

\begin{figure}
\centering
 \plotone{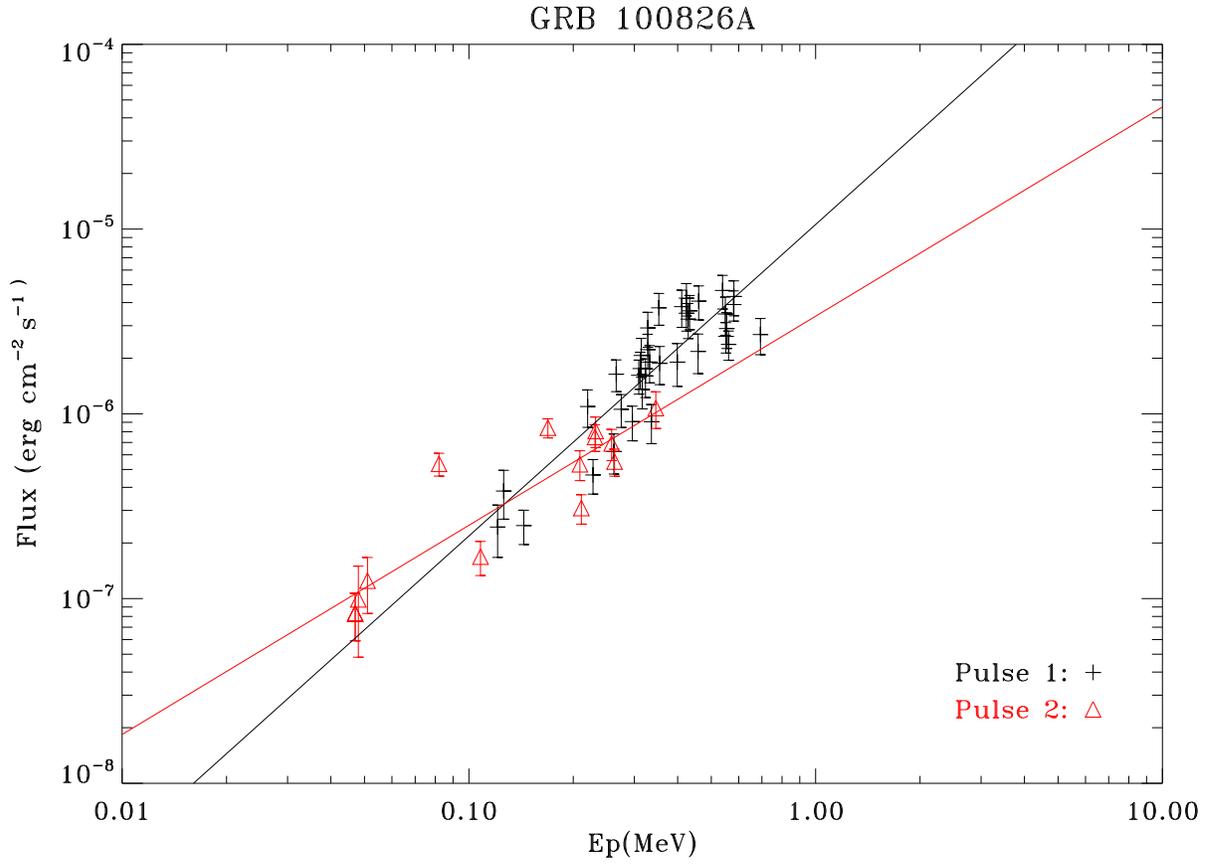}
  \caption{The  plot of $F_p\sim E_p$ relation in GRB 100826A. The fitting  lines  are given by ${\rm log}F_p=-4.975 \pm 0.077 +(1.686\pm0.158){\rm log}E_p$ and ${\rm log}F_p=-5.471 \pm 0.151 +(1.132\pm0.162){\rm log}E_p$, respectively. }  \label{fig:100826flux}
\end{figure}

\begin{figure}
\centering
 \plottwo{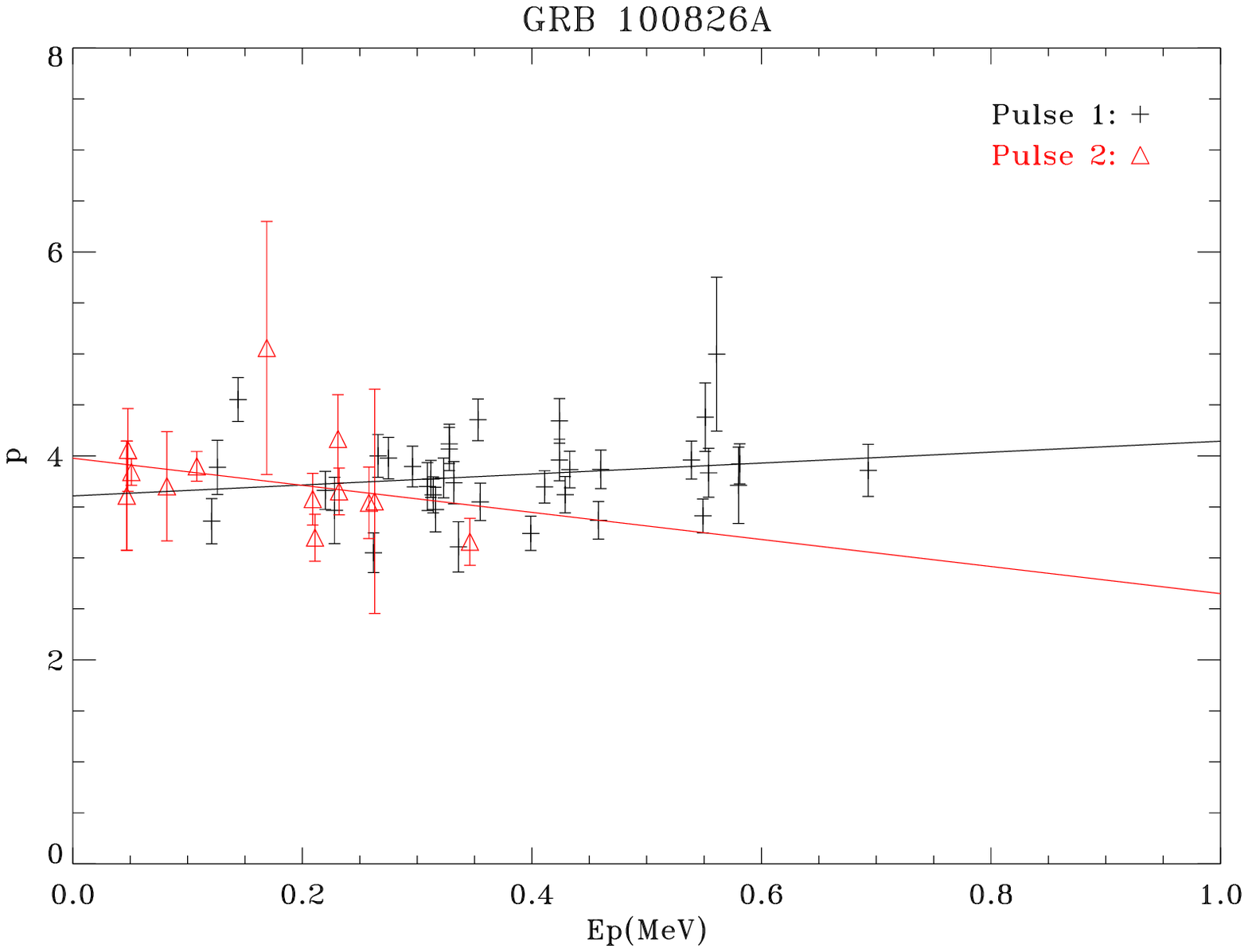}{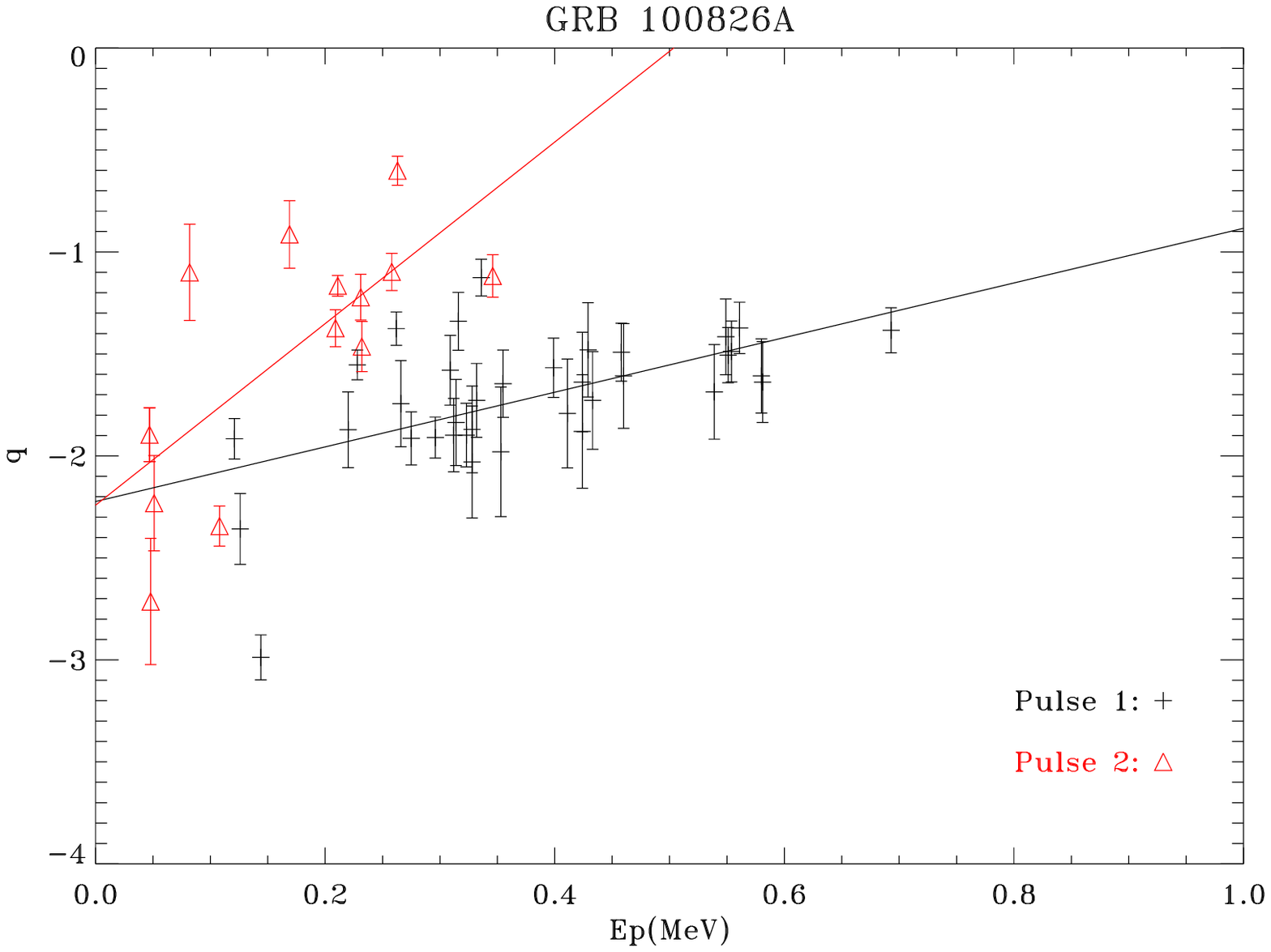}
  \caption{The plots of $p$ and $q $ versus $E_p$ relation in GRB 100826A. In the left panel, the line fittings of $p$ versus $E_p$ are described by
   $p=3.608 \pm 0.196 +(0.536 \pm 0.489)E_p$ and $3.976 \pm 0.247 -(1.327\pm 1.298)E_p$ in pulse 1 and 2, respectively. In the right panel, the
   linear fitting of $q$ are given by $q=-2.223\pm 0.135 +(1.339\pm0.335)E_p$ and $q=-2.242\pm 0.233+(4.45\pm 1.224)E_p$ in pulse 1 and 2, respectively. } \label{fig:100826p}
\end{figure}

\begin{figure}
\centering
 \plottwo{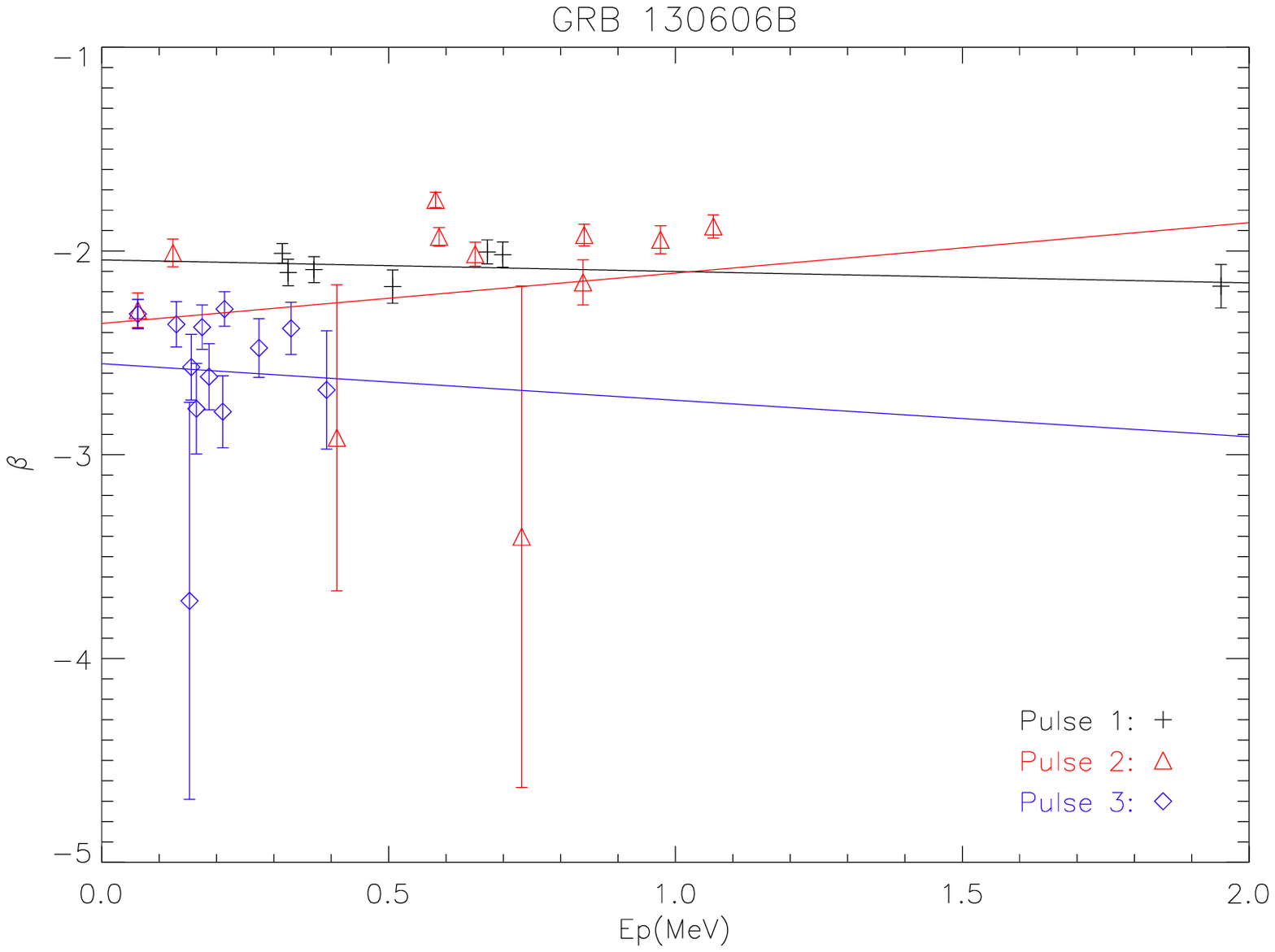}{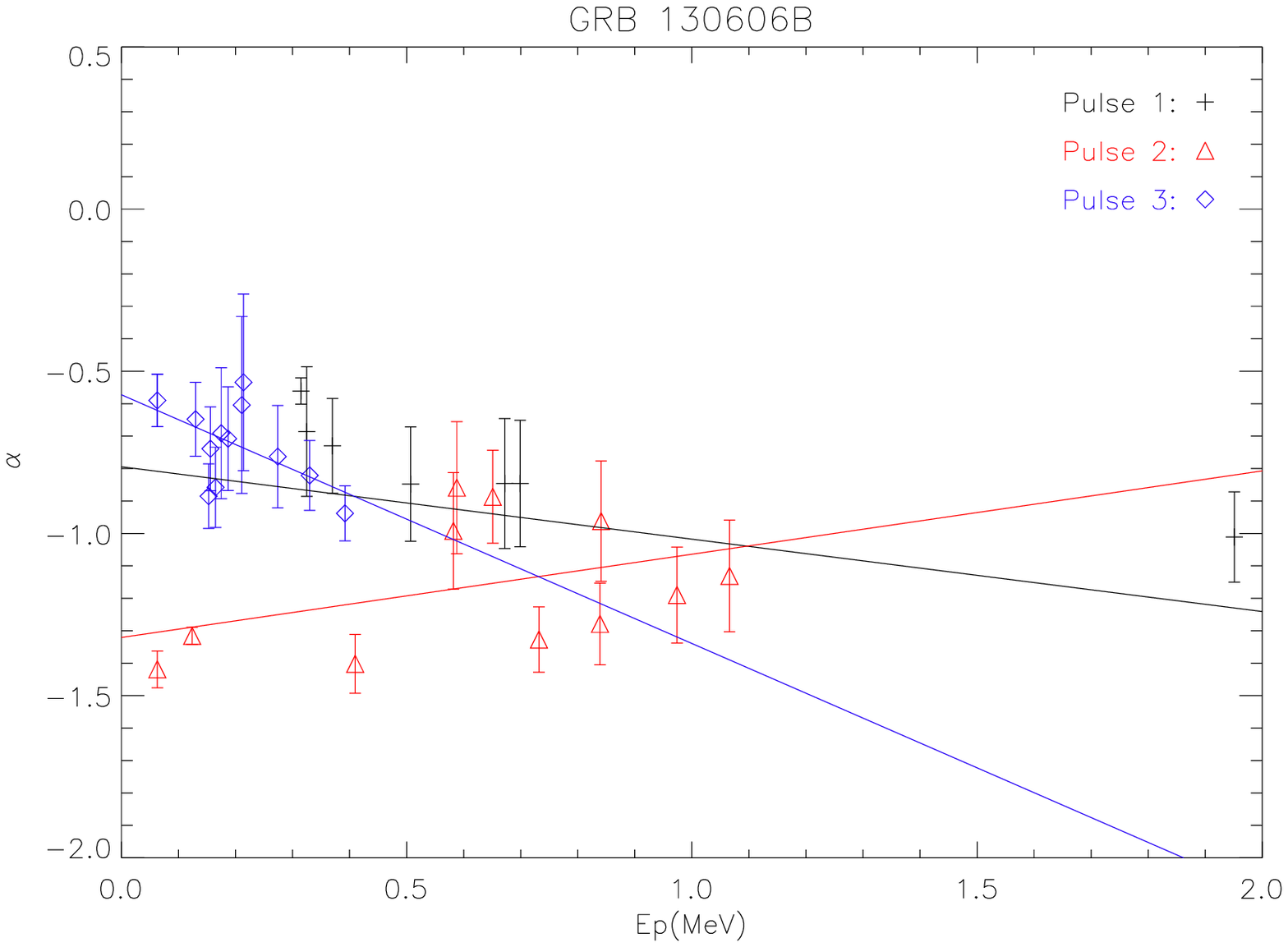}
  \caption{The  dependence of $\beta$ and $\alpha$ on $ E_p$  in GRB 130606B. In the left panel, the three fitting lines in pulse 1, 2, and 3
  are $\beta=-2.044\pm0.044-(0.056\pm0.051)E_p$, $\beta=-2.355\pm0.360+(0.247\pm0.517)E_p$, and $\beta=-2.55\pm0.260-(0.179\pm1.21)E_p$, respectively. In the right panel, the fitting lines of $\alpha$ are given by  $-0.794\pm0.068-(0.223\pm0.112)E_p$, $-1.32\pm0.136+(0.257\pm0.196)E_p$, and $-0.572\pm0.071-(0.767\pm0.333)E_p$ in pulse 1, 2, and 3, respectively. }  \label{fig:130606ba}
\end{figure}

\begin{figure}
\centering
 \plotone{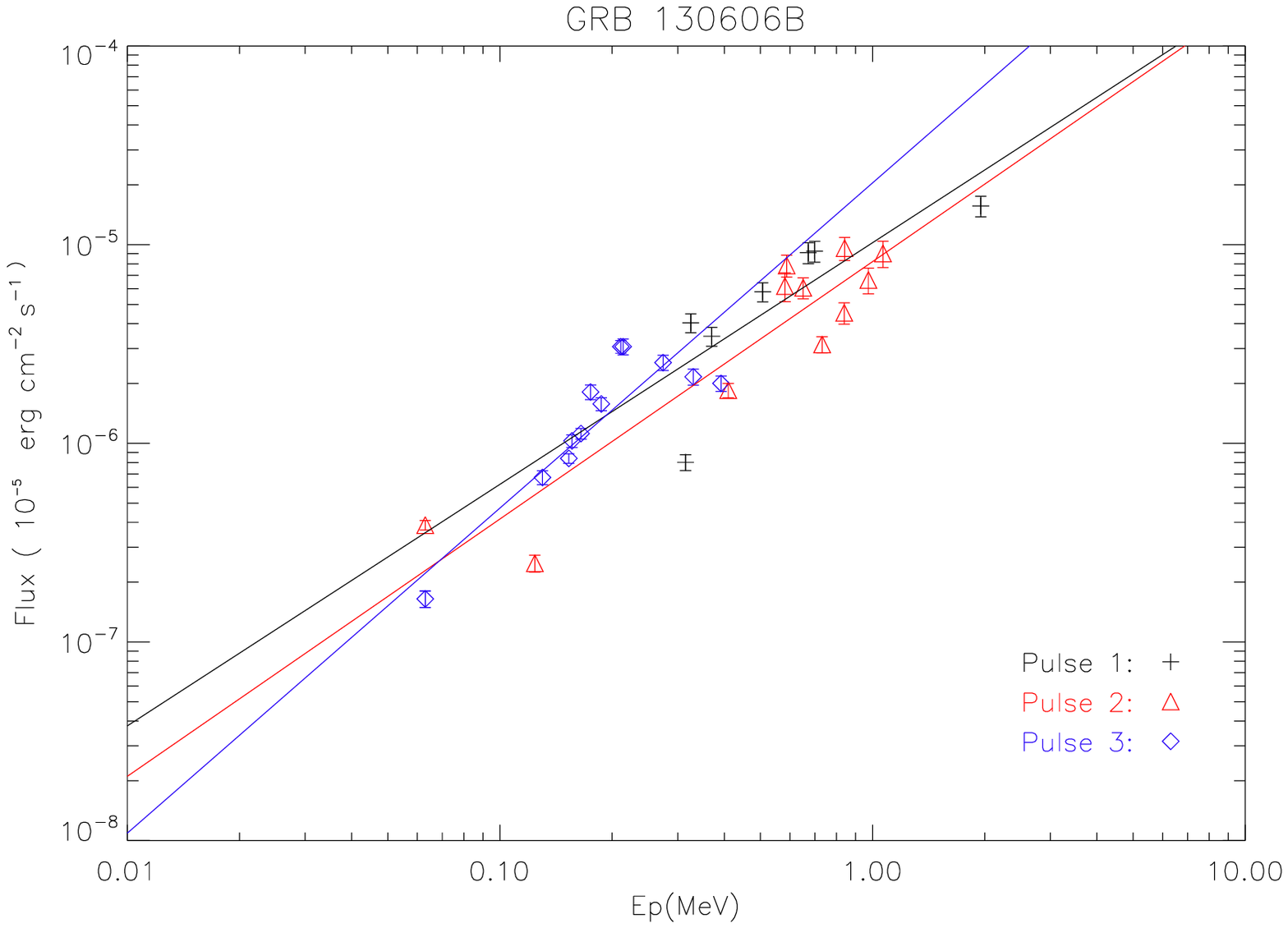}
  \caption{The  plot of flux on $\sim E_p$  in GRB 130606B. The three fitting lines are described by ${\rm log}F_p= -4.990\pm0.144+(1.216\pm0.401) { \rm log}E_p$,
   ${\rm log}F_p= -5.085\pm0.086+(1.293\pm0.177){\rm log}E_p$, and ${\rm log}F_p= -4.689\pm0.183+(1.636\pm0.229){\rm log}E_p$ in pules 1, 2, and 3, respectively. }L \label{fig:130606flux}
\end{figure}

\begin{acknowledgments}
This work has been funded by the National Natural Science Foundation of China under Grant No. 11403015, No.11203016 and No.11143012. This work is partly supported by the Natural Science Foundation of Shandong Province under grant No. ZR2012AQ008 and No. ZR2014AQ007.
\end{acknowledgments}

\end{document}